\documentclass[aps,prl,twocolumn,superscriptaddress]{revtex4}
\usepackage{graphicx}
\usepackage{latexsym}
\usepackage{amssymb}
\usepackage{amsmath}
\usepackage{amsfonts}
\usepackage{bm}
\usepackage{multirow}
\usepackage{color}
\newcommand{\ii}{\mathrm{i}}

\newcommand{\ket}[1]{|#1 \rangle}
\newcommand{\dsZ}{\mathbb{Z}}
\newcommand{\dsR}{\mathbb{R}}
\newcommand{\dsC}{\mathbb{C}}
\newcommand{\dsH}{\mathbb{H}}

\renewcommand{\mod}{\mathop{\mathrm{mod}}}

\newcommand{\SU}{\mathrm{SU}}
\newcommand{\U}{\mathrm{U}}

\newcommand{\Cl}{\mathcal{C}\ell}
\newcommand{\vect}[1]{{\bm{#1}}}
\newcommand{\eqnref}[1]{Eq.\,\eqref{#1}}
\newcommand{\figref}[1]{Fig.\,\ref{#1}}
\newcommand{\tabref}[1]{Tab.\,\ref{#1}}

\newcommand{\mat}[1]{\left(\begin{smallmatrix}#1\end{smallmatrix}\right)}
\newcommand{\eq}[1]{\begin{equation}#1\end{equation}}
\newcommand{\eqs}[1]{\begin{equation}\begin{split}#1\end{split}\end{equation}}

\begin{document}

\title{Sachdev-Ye-Kitaev Model and Thermalization on the Boundary of Many-Body Localized Fermionic Symmetry Protected Topological States}

\author{Yi-Zhuang You}
\author{Andreas W. W. Ludwig}
\author{Cenke Xu}

\affiliation{Department of physics, University of California, Santa Barbara, CA 93106, USA}  

\date{\today}
\begin{abstract}
We consider the Sachdev-Ye-Kitaev (SYK) model\cite{Sachdev:1993lr,Sachdev:2015ao,Kitaev:2015uj} as an effective theory arising 
at the 
zero-dimensional  boundary of 
a many-body localized, Fermionic symmetry protected topological (SPT) phase in one spatial dimension. 
The Fermions 
at
the boundary are always fully interacting.  
We find that  the boundary is  thermalized and investigate how  its boundary anomaly, dictated by the bulk SPT order,  is encoded in the  quantum chaotic eigenspectrum of the SYK model. We show that depending on the SPT symmetry class, 
the boundary many-body level statistics 
cycle in a systematic manner through those of the three different Wigner-Dyson 
random matrix ensembles with a periodicity in the topological  index that matches the 
interaction-reduced classification of the bulk SPT states.  We consider all three symmetry classes BDI, AIII, and CII, whose SPT phases are 
classified  in  one spatial dimension
by $\dsZ$
in the absence of interactions.
For symmetry class BDI, we derive the eight-fold  periodicity of the Wigner-Dyson statistics by using Clifford algebras. 
\end{abstract}

\maketitle  

{\it Introduction.} Symmetry protected topological (SPT) phases are gapped quantum systems with quantum disordered short-range
entangled ground states which cannot be smoothly deformed into trivial product states without closing the gap, if the symmetry defining the SPT phase  is preserved.
The corresponding ground states
are non-degenerate even on spatial manifolds with non-trivial topology.\cite{Wen:2009qv,Wen:2011sf, Fidkowski:2011pa} 
Famous examples of SPT states include the ground states of the Haldane spin-1 
chains\cite{Haldane:1983fu,Haldane:1983vl,AKLT:1987dp} and of   topological insulators and  superconductors\cite{Kitaev:2001un,Read:2000sc,Kane:2005lr,Moore:2007ao,Hasan:2010rv,Qi:2011rv,Ludwig:2008zy,Ludwig:2010xt,Kitaev:2009sd,Ludwig:2015la}.  Short-range entanglement is manifested in  a strict area-law  entanglement entropy of the SPT state. As a quantum order, the SPT order is in general  not expected to persist to highly-excited (finite-energy-density) states in the many-body spectrum, because highly excited states are typically thermalized  according to the eigenstate thermalization hypothesis (ETH),\cite{Deutsch:1991ik,Srednicki:1994ns,Rigol:2008qr,Rigol:2012wd} and have a volume-law entanglement entropy in contrast to the area-law entanglement in the SPT state.  

However the phenomenon of many-body localization (MBL)\cite{BAA:2006,GMP:2005,Huse:2007is,Znidaric:2008he,Imbrie:2014,Huse:2015rv} provides a class of examples where quantum many-body systems can evade thermalization in the presence of quenched disorder. MBL systems are generic non-ergodic phases of matter, which can retain the memory of local quantum information and exhibit  an area-law entanglement entropy even for highly excited (finite energy-density) states. In some sense,  excited states of an MBL system are like ground states\cite{Bauer:2013al}, which enables us to extend the discussion of ground state quantum orders to highly excited (finite energy-density) states.  Examples of  many-body localization protected quantum order have recently been discussed in Ref.\,\cite{Huse:2013ci,Bauer:2013al,Vishwanath:2013tp,Altman:2014hg,Chandran:2014fv,Altman:2014dq,Vishwanath:2015po,Xu:2015gg}. Here we are interested in  such  ``MBL stabilized SPT states'' (referred to hereafter as ``MBL-SPT states'').  

In particular, we will focus on  interacting Fermionic MBL-SPT states, and investigate the possibility and the consequences of thermalization at the boundary of such a state. Many-body localizability of SPT states has been discussed in Ref.\,\cite{Vishwanath:2015po, Xu:2015gg}. It was shown that, at least in one spatial  dimension (1D), Fermionic SPT states can be fully many-body localized in the bulk. In the strong disorder  regime of the MBL system,  all bulk Fermion degrees of freedom can be renormalized to local integrals of motions (LIOMs).\cite{Abanin:2013ta,Abanin:2013lc,Huse:2014ec,Scardicchio:2015kl,Abanin:2015io,Rademaker:2015ve,You:2015sb} Resonances among LIOMs are suppressed by disorder, which makes the bulk stable against thermalization. However, the Fermions near the boundary of the MLB-SPT system are less protected against thermalization. Indeed, if the 1D bulk SPT order is non-trivial, its zero-dimensional (0D) boundary will host degenerate boundary  (``edge'')  states (generalizing the Majorana Fermion zero modes in the non-interacting limit),\cite{Huse:2013ci}  whose presence reflects a quantum anomaly\cite{RyuMooreLudwig:2012PRB,Wen:2013qf,RyuNobelSymp2015,LudwigNobelSymp2015,WittenTopPhases2015} that is required by the SPT order in the bulk. For example, consider the symmetry class\cite{AZclass,HeinznerCMP2005,Ludwig:2010xt,Ludwig:2008zy} BDI, whose 1D non-interacting Fermion SPT phases are classified by  $\dsZ$\cite{Kitaev:2009sd,Ludwig:2008zy,Ludwig:2010xt};
 its 0D  boundary can then support arbitrary many Fermion zero modes in the absence of interactions.  Due to the vanishing level-spacing among the edge states, the boundary can be easily thermalized once   interactions are introduced to couple the Fermion edge modes together. Then we are facing  the interesting scenario of  an  MBL-SPT bulk with thermalized boundaries. One may wonder if the MBL bulk is stable against thermalization due to the contact with the thermalized boundaries. According 
renormalization group (RG) studies of the  MBL-ETH transition in one dimension\cite{Huse:2014cs,Potter:2015kl}, the MBL fixed-point is stable against weak thermalization:  A thermal bubble (small thermalized region) in the  1D bulk cannot expand indefinitely into the MBL environment. Therefore, the thermalized  boundary (which can be viewed as  a thermalized bubble residing at the boundary) will not be able to thermalize the entire bulk in 1D.  

If the  effective Hamiltonian of the boundary  is in the ETH phase, we say that the boundary is thermalized. However, if the MBL bulk has  non-trivial SPT order, the ETH boundary must  possess a  corresponding anomaly that characterizes the SPT phase\cite{Fidkowski:2011pa,RyuMooreLudwig:2012PRB,Wen:2013qf,RyuNobelSymp2015,LudwigNobelSymp2015}.  What is the signature of this quantum anomaly for a thermalized boundary?  First of all,  the presence of a   protected  degeneracy of every energy level in the boundary many-body spectrum\cite{Huse:2013ci,Vishwanath:2013tp} is one obvious signature. In this work, we will show that the \emph{level statistics} of the boundary spectrum is another such signature. In particular we will show that for the thermalized boundary, the level statistics follows the Wigner-Dyson (WD) distribution  of one of the three  WD random matrix ensembles  which is in correspondence, as specified below, with  the global anomaly required by the bulk SPT order: Take for example the thermalized boundary of the MBL-SPT state in symmetry class BDI. We find that its level statistics
cycles through that of
 the Gaussian orthogonal, unitary, and symplectic ensembles (GOE, GUE, and GSE) 
in a systematic manner,
as summarized in \tabref{tab: BDI} with an \emph{eight-fold periodicity} that matches the (interaction-reduced) $\dsZ_8$ classification\cite{Fidkowski:2010iv,Fidkowski:2011pa}  of the  Fermionic SPT order in symmetry class BDI.

Subsequently,  we will extend our analysis to the boundary level statistics for Fermionic MBL-SPT states in  symmetry classes AIII and CII, which are the other two 1D symmetry classes that also possess a $\dsZ$ classification in the absence of
interactions.\cite{Kitaev:2009sd,Ludwig:2008zy,Ludwig:2010xt}
In contrast to symmetry class BDI discussed above, here we have to pay attention to the fact that Fermionic MBL-SPT states in 
classes AIII and CII are in general unstable to interactions in the 1D 
bulk\cite{Potter:2016uq} (due to the presence of charge-conjugation symmetry).\footnote{We thank Andrew Potter for drawing our attention to this fact.}
For this reason, we will consider in symmetry classes AIII and CII situations where interactions are solely present at the 0D boundaries, 
whereas
the 1D bulk remains non-interacting throughout. The resulting 0D boundaries also turn out to be precisely described by Sachdev-Ye-Kitaev
models \cite{Kitaev:2015uj,Sachdev:1993lr,Sachdev:2015ao} (or corresponding generalizations). The extremely rich physics of SYK models are being actively explored recently.\cite{Maldacena:2016tg,Polchinski:2016qq,Gu:2016wq,Banerjee:2016dk,Witten:2016pb,Fu:2016la,Berkooz:2016th,Gross:2016fy} Even though in symmetry classes AIII and CII
these bulk states are not many-body localizable (when bulk interactions are turned on), we can (and will) nevertheless still discuss the
spectral properties of corresponding interacting boundary Hamiltonians, which turn out to be, as mentioned, the corresponding SYK models (while always considering
a corresponding non-interacting, but random bulk).
- We finally note that   our discussion of level statistics also applies to bosonic MBL-SPT states, because bosonic SPT states in 1D can always be interpreted as interacting Fermionic SPT states.\cite{You:2014pt,You:2015bx}

Finally, as a byproduct of our analysis we show that  in general, anti-unitary operators such a time-reversal or Chiral symmetry operation
 acting on the many-body Fermionic Fock space  in any spatial dimension
square to the  conventional form $(\pm 1)^F$, times  a ``{\it many-body phase}'' which is a fourth root of
unity.\cite{Fidkowski:2013fk,Metlitski:2014fv} The precise form of this statement, Eq.s (\ref{eq: T^2 SP},\ref{eq: T^2 MB},\ref{LabelEqDEFTimeReversalSquareComplexFermions},\ref{LabelEqDEFChiralSquareComplexFermions},\ref{eq: dep}), as well as
its proof, are presented in the Appendix.


\begin{table}[htbp]
\caption{Eight-fold-way spectrum on the thermalized boundary  of $N_\chi$ Majorana chains in symmetry class  BDI  ($N_\chi>4$). qdim: quantum dimension (level degeneracy per boundary), lev.\,stat.: level statistics in a definite Fermion number parity sector.}  
\begin{center}
\begin{tabular}{c|cccccccc}
$N_\chi(\mod 8)$ & 0 & 1 & 2 & 3 & 4 & 5 & 6 & 7\\
\hline  
qdim & $1$ & $\sqrt{2}$ & $2$ & $2\sqrt{2}$ & $2$ & $2\sqrt{2}$ & $2$ & $\sqrt{2}$ \\
lev.\,stat. &  GOE & GOE & GUE & GSE & GSE & GSE & GUE & GOE\\
$\Cl_{0,N_\chi-1}$ & $\dsR\oplus\dsR$ & $\dsR$ & $\dsC$ & $\dsH$ & $\dsH\oplus\dsH$ & $\dsH$ & $\dsC$ & $\dsR$
\end{tabular}
\end{center}
\label{tab: BDI}
\end{table}

\vskip .1cm 

{\it Symmetry class BDI, four fermi interactions, numerical results.} We will start with 1D Fermionic MBL-SPT states in symmetry class  BDI, sometimes also known as the ``Kitaev Majorana wire'',\cite{Kitaev:2001un} protected by a time-reversal symmetry $Z_2^T$ which squares to the identity operator (in the single-particle setting - for a complete discussion of the action of the  square of the time-reversal operator on the many-body Fock space, see \cite{SuppII}). 
In the absence of interactions, 
the SPT order is characterized by  an integer-valued topological index $N_\chi\in\dsZ$, which counts the number $N_\chi$ (of species) of protected Majorana zero modes $\chi_a$ ($a=1,2,\cdots,N_\chi$) at the boundary. The operators $\chi_a$  satisfy the Clifford algebra $\{\chi_a,\chi_b\}=2\delta_{ab}$ and $\chi_a^\dagger =\chi_a$.  The (anti-unitary)  time-reversal symmetry acts on these Majorana zero modes as $\mathcal{T}\chi_a\mathcal{T}^{-1} = \chi_a$.  Fermion bilinear terms $\ii\chi_a\chi_b$ are forbidden to occur in the boundary Hamiltonian $H$  by  time-reversal symmetry $\mathcal{T}$. So to  lowest order in  many-body terms, the boundary dynamics  is governed by  random four-Fermion interactions  
\eq{\label{eq: Kitaev} H = \sum_{a<b<c<d}V_{abcd}\chi_a\chi_b\chi_c\chi_d.}
Here,  the  interaction strengths $V_{abcd}$ are  taken to be independent random real numbers with zero mean. The randomness in the boundary Hamiltonian, \eqnref{eq: Kitaev}, 
originates from the strong disorder in the 1D MBL bulk. The detailed probability distribution of $V_{abcd}$ is unimportant, and  we may assume it to be Gaussian. This  model, \eqnref{eq: Kitaev},  was introduced by Kitaev\cite{Kitaev:2015uj} as a toy model for holography. Here we would like to consider it as an effective model describing the boundary of  a 1D SPT phase.  From this perspective, the fact that this model contains no Fermion bilinear term (a condition imposed by hand in Kitaev's model) appears here
naturally 
as a consequence of the symmetry requirement (in the present case,  the relevant symmetry is the  time-reversal symmetry in class BDI).  

In general, the 
 degeneracy at the boundary of the non-interacting system arising from the Majorana zero modes (which is vast when $N_f$
is large)  can be lifted by interactions. However,  if the bulk SPT order is non-trivial, the boundary degeneracy cannot be fully lifted, because otherwise the bulk state could have been smoothly deformed into the trivial vacuum state across the boundary (which here is a boundary
to vacuum).  We recall that with interactions, the classification of the  SPT order in symmetry class BDI  is reduced from $\dsZ$ to $\dsZ_8$.\cite{Fidkowski:2010iv,Fidkowski:2011pa} So the energy levels  of the boundary many-body spectrum are non-degenerate if and only if $N_\chi$ is a multiple of eight [$N_\chi (\mod 8) =0$]; otherwise, there is a degeneracy
of every energy level of the many-body spectrum. The 
degeneracy of
energy levels of the boundary many-body Hamiltonian
 can be  studied  numerically by exact  diagonalization of the  Hamiltonian in \eqnref{eq: Kitaev}. In doing so we need to recall that when $N_\chi$ is odd, the low-energy Hilbert space of a single boundary is not well-defined. In that case, the ``quantum dimension''  (${\rm qdim}$) of the boundary mode is considered instead, which is defined to be  the square root of the level degeneracy with both boundaries considered. Numerical results for ${\rm qdim}$  are  listed  on the second line of \tabref{tab: BDI}: We see that the  eight-fold periodicity of the level degeneracy matches the $\dsZ_8$  periodicity of the (global) anomaly  of  the boundary.  

However the level degeneracy (or ``quantum dimension'') alone cannot fully  resolve the eight-fold anomaly described by $\dsZ_8$.  
As we will now explain, we
have  found  that the level statistics can provide an additional diagnostic. In the past, the level statistics of the many-body spectrum has been used to diagnose whether a many-body Hamiltonian is in the MBL phase or the ETH phase (see e.g. \cite{Huse:2007is,Rigol:2010ls}). Here we use the level statistics of the boundary to further resolve the (global) quantum anomaly of the SPT phase, beyond the  diagnostic
provided by the degeneracy of all levels.  In general, we collect the eigen energies $\{E_n\}$ of the Hamiltonian, and arrange them in ascending order $E_1<E_2<\cdots$. Let $\Delta E_n=E_n-E_{n+1}$ be the 
level spacing, and we evaluate the ratios of adjacent level spacings $r_n=\Delta E_n/\Delta E_{n+1}$,\cite{Huse:2007is,Chandran:2014ls,Reichman:2015sf} such that the  dependence on the density of states cancels out in the ratio. The distribution of the ratio $r_n$ follows  Poisson level statistics in the MBL phase,  
\eq{\label{eq: r-stat MBL} \text{Poisson: }p(r) = \frac{1}{(1+r)^2},}
 and  WD level statistics in the ETH phase (given by the ``Wigner-surmise''\cite{Atas:2013la}),  
\eq{\label{eq: r-stat ETH} \text{WD (surmise): }p(r) = \frac{1}{Z}\frac{(r+r^2)^\beta}{(1+r+r^2)^{1+3\beta/2}}.}
 The parameters $\beta$ and $Z$ are different for GOE: $\beta=1, Z=\tfrac{8}{27}$; GUE: $\beta=2, Z=\tfrac{4\pi}{81\sqrt{3}}$; and GSE: $\beta=4, Z=\tfrac{4\pi}{729\sqrt{3}}$. The level repulsion in the ETH spectrum  manifests itself  in the asymptotic behavior $p_\text{WD}(r\to0)\sim r^\beta$. To make clearer the contrast  between different level statistics, we choose to show the  probability distribution of the logarithmic ratio $\ln r$,  which is given by $P(\ln r)=p(r) r$.  

\begin{figure}[htbp]
\begin{center}
\includegraphics[width=200pt]{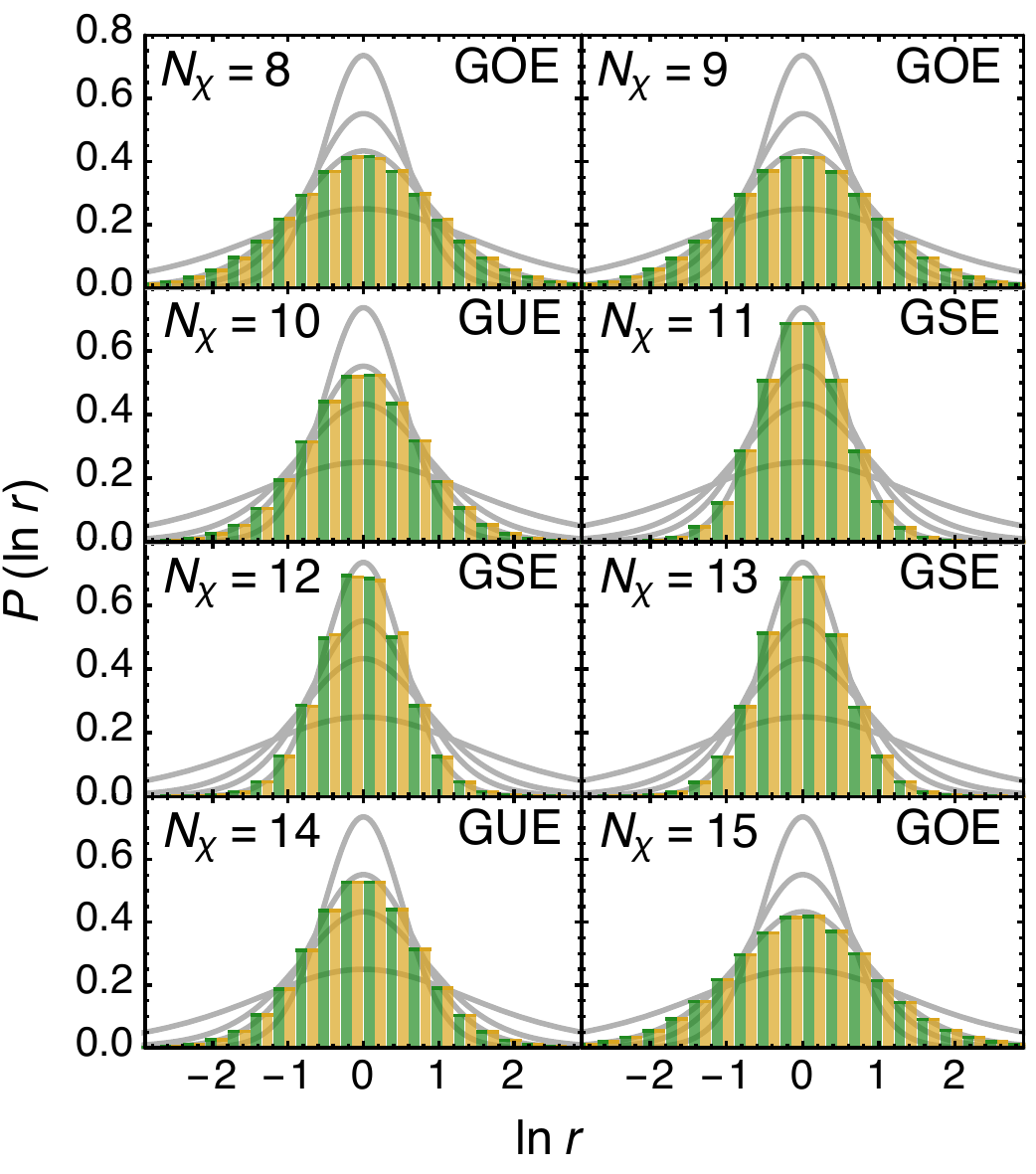}
\caption{Many-body level statistics (in term of the $\ln r$ distribution) of the random interaction model in \eqnref{eq: Kitaev}, for $N_\chi=8,\cdots,15$ (a full $\dsZ_8$ period) by exact diagonalization. The background gray curves describe  the ``Wigner-surmise'' given by \eqnref{eq: r-stat MBL} and \eqnref{eq: r-stat ETH}: from wide to narrow, they correspond to Poisson, GOE, GUE, and GSE statistics respectively. The level statistics in the even (odd) Fermion parity sector is shown in green (yellow).} \label{fig: level statistics}  
\end{center}
\end{figure}

We now apply  this analysis to the boundary Hamiltonian in \eqnref{eq: Kitaev}. However, extra care should be taken regarding the Fermion parity. Levels with different Fermion parities are independent, so putting all levels together will spoil the true level statistics in each sector.\footnote{If the spectrum is not separated by symmetry sectors, the level statistics will typically be Poissonian even if the system is in the ETH phase, because there is no level repulsion between different symmetry sectors.} Therefore, the level statistics must be collected in each Fermion parity sector. 
Since  our BDI-class Hamiltonian in  \eqnref{eq: Kitaev}
possesses, besides Fermion number parity, 
no  other unitary symmetries,
any remaining level degeneracies within each Fermion parity sector will be ignored, i.e.\;we only consider the level spacing between  adjacent (non-degenerate) eigenenergies in each such sector.
 We have collected the probability distribution $P(\ln r)=p(r)r$ of the logarithmic ratio $\ln r$ numerically; the results are  shown in \figref{fig: level statistics}. We see that the probability distribution   varies systematically with the number $N_\chi$ of Majorana modes. First of all,  in all cases WD statistics is observed, which shows that the boundary is indeed in the ETH (quantum chaotic) phase. Secondly, depending on the topological index $\nu\equiv N_\chi (\mod 8$), the data correspond to one of the three  WD random matrix ensembles (GOE, GUE, or GSE),  as summarized on  the third line of \tabref{tab: BDI}.\footnote{$N_\chi$ must be sufficiently large ($N_\chi>4$), in order to observe the universal WD level statistics. If $N_\chi$ is too small, the system is not thermalized, and hence no WD statistics is observed.} Combining the results for the  level statistics (3rd line of \tabref{tab: BDI})   with  those for  the level degeneracy  (2nd  line of \tabref{tab: BDI}) , the $\dsZ_8$ anomaly pattern of  the thermalized boundary can be determined up to the sign of the topological index $\nu$ (i.e. $\nu$ and $-\nu$ are not distinguishable yet).\footnote{$\nu$ and $-\nu$ may be further distinguished by the projective representation of the time-reversal and Fermion parity combined symmetry $P\mathcal{T}$. For more details, see the Appendix \ref{sec: BDI}.}

\vskip .1cm 

{\it Symmetry class BDI, most general Hamiltonian, analytical results.} In this section we demonstrate analytically that the `eight-fold-way'  level statistics of the boundary Hamiltonian \eqnref{eq: Kitaev} persists even after including all possible (random) higher-order interactions
(see Eq. (\ref{eq: H BDI}) below).
Moreover, we show that this   is related to the Bott periodicity of the real Clifford algebra $\Cl_{0,N_\chi-1}$.\footnote{$\Cl_{p,q}$ denotes the real Clifford algebra with $p$ symmetric generators and $q$ antisymmetric generators.} To make this connection, let us first  observe that the  Fermion bilinear operators  
\eq{ 
\label{Clifford}
\gamma_{a}=\chi_{a}\chi_{N_\chi}\quad(a=1,2,\cdots,N_\chi-1),}
where  $\chi_{N_\chi}$ is the ``last'' of the $N_\chi$ Majorana modes on the boundary,  can be used to define the generators of
the Clifford algebra $\Cl_{0,N_\chi-1}$. We consider a real (matrix)  representation in  Fock space, so that  we have $\chi_a^\intercal =\chi_a$  (where ${}^\intercal$ denotes the transposed matrix), and $\{\chi_a,\chi_b\}=2\delta_{ab}$. Then it is easy to show, using \eqnref{Clifford},   that $\gamma_a^\intercal =-\gamma_a$ and $\{\gamma_a,\gamma_b\}=-2\delta_{ab}$. So  the operators  $\gamma_a$ indeed represent the $(N_\chi-1)$ antisymmetric generators of the Clifford algebra $\Cl_{0,N_\chi-1}$. Then it can be checked that those elements in $\Cl_{0,N_\chi-1}$ which are represented by symmetric matrices (in the real representation we are currently considering -- they are thus self-adjoint) are of grade $(4k-1)$ or $4k$ (for some $k\in\dsZ_+$), meaning that they can be written as products of $(4k-1)$ or $4k$ generators $\gamma_{a}$. It turns out that these  matrices represent all  possible  time-reversal invariant  terms that are allowed in the boundary Hamiltonian. For example, the four Fermion interaction terms in \eqnref{eq: Kitaev} are of grade 3 and 4 (corresponding to $k=1$): $\chi_{a}\chi_{b}\chi_{c}\chi_{N_\chi}=-\gamma_{a}\gamma_{b}\gamma_{c}$ ($a,b,c < N_\chi$) are grade-3 terms, and $\chi_{a}\chi_{b}\chi_{c}\chi_{d}=\gamma_{a}\gamma_{b}\gamma_{c}\gamma_{d}$ ($a,b,c,d < N_\chi$) are grade-4 terms. Higher order time-reversal invariant interactions ($4k$-Fermion interactions) correspond to higher grades in the Clifford algebra $\Cl_{0,N_\chi-1}$, and it is not difficult to see that these exhaust the full space of all  symmetric matrices  in $\Cl_{0,N_\chi-1}$ (in a real representation).  Therefore, if all symmetry-allowed interactions are included in the Hamiltonian,  
\eq{\label{eq: H BDI} H=\sum_{k=1}^{\lfloor N_\chi/4\rfloor}\sum_{a_1<\cdots<a_{4k}}V_{\{a_i\}} \ \chi_{a_1}\cdots\chi_{a_{4k}},}
the Hamiltonian $H$ will be a general (real)   symmetric matrix in (the real representation of) the Clifford
algebra
 $\Cl_{0,N_\chi-1}$. Hence, when the real coefficients $V_{\{a\}}$ are random, $H$ will be a general  random  symmetric matrix in $\Cl_{0,N_\chi-1}$, and its  level statistics will fall into the matrix ensemble 
determined by the real  representation of $\Cl_{0,N_\chi-1}$. The representations of 
the Clifford algebra
$\Cl_{0,N_\chi-1}$ are known and are listed in the last line of \tabref{tab: BDI}, where $\dsR$, $\dsC$ and $\dsH$ stand respectively for the set of all $m\times m$ matrices with  real, complex and quaternion entries with certain matrix dimensions $m$, which are not written out explicitly; 
the Hermitian such matrices (symmetric in a real representation of the corresponding Clifford algebra)
correspond, respectively,  to the Hamiltonians in the three Wigner-Dyson  random matrix ensembles GOE, GUE and GSE.  The 
numerical results  reported in the previous sections of this paper for a Hamiltonian of the form of  \eqnref{eq: H BDI}, which contains solely four-Fermion interactions,  show that restricting the generic Hamiltonian in \eqnref{eq: H BDI} to one that contains only four-Fermion interactions (as  in \eqnref{eq: Kitaev}),  does not affect the level statistics.  In fact, all  higher order interactions allowed by symmetry can  be generated under  the renormalization group, 
so the properties of the Hamiltonians in 
 \eqnref{eq: Kitaev} and \eqnref{eq: H BDI} 
are indeed not expected to be fundamentally different.

In the remainder of this paper we extend the above discussion for the  MBL-SPT state in symmetry class BDI to the other two
 symmetry classes that also have  a $\dsZ$ classification in 1D in the absence of interactions:  These are symmetry classes AIII and CII which we will now discuss in turn. 
As already mentioned in the introduction, 
here we need to pay attention to the fact that, in contrast to symmetry class BDI, there are no
many-body localized 
Fermionic SPT phases  in symmetry classes AIII and CII in the 1D bulk.\footnote{Following the arguments in Ref. \onlinecite{Potter:2016uq},
the symmetry group in these two classes admit, owing to the presence of charge conjugation symmetry,
 irreducible representations of dimensions larger than one, which turn out to imply an extensive number of local degeneracies for any SPT-MBL
state which are  invariant under the protecting symmetry group. This  extensive degeneracy  makes the system unstable
in the presence of interactions. This leads  to either a spontaneous breaking of the symmetry protecting the SPT order, or thermalization.
All these scenarios destroy the boundary modes that would otherwise be a necessary consequence
of SPT order.} For this reason, in symmetry classes AIII and CII we consider situations in which interactions are only present at the 0D boundaries,
the 1D bulk system remaining throughout a non-interacting Anderson-localized insulator. The non-interacting Anderson-localized 1D {\it bulk}
systems in these two symmetry classes
are known to possess a $\dsZ$ classification which applies\cite{Bauer:2013al} also to excited states (at finite energy density),
in complete analogy  with the usual MBL systems  (in which interactions are present). 
Here  the interactions, localized solely at the boundary, reduce  the
$\dsZ$ classification of the 0D boundary Hamiltonian to  
$\dsZ_4$ in class AIII, and to $\dsZ_2$ in class CII. The following analyses of the level statistics (which are confirmed numerically)
 of the resulting SYK systems
at the 0D boundaries in symmetry classes AIII and CII  must then follow
as long as the respective
protecting symmetries are not broken spontaneously in (0+1) dimensions.


\vskip .1cm 

{\it Symmetry Class AIII.}
Ground states of 1D
 SPT  phases  in symmetry class AIII 
can be viewed as being protected by  $\U(1)\times Z_2^{\mathcal{S}}$ symmetry, where the superscript ${}^{\mathcal{S}}$ stands
for {\it  chiral} symmetry $\mathcal{S}$.\cite{AZclass,HeinznerCMP2005,Ludwig:2008zy,Ludwig:2010xt,Ludwig:2015la} 
The 1D  SPT phases protected by this symmetry are classified by  $\dsZ_4$  in the presence of interactions. 
In the sense explained above, the resulting boundary anomaly will determine the properties of all states of the random boundary
Hamiltonian in this class.
The corresponding  0D  boundary degrees of freedom are complex Fermion modes $c_a$ ($a=1,2,\cdots,N_c$), where  $N_c$ labels the number of complex Fermion mode species. The $\U(1)$ symmetry is naturally implemented as $c_{a}\to e^{\ii\theta}c_a$. The
{\it chiral} symmetry, an  anti-unitary symmetry operation when acting on the many-body Fermion Fock space, 
can be taken to
act as 
$
\mathcal{S}c_a\mathcal{S}^{-1}=c^\dagger_a,  \quad
{\rm and} \ 
\mathcal{S}c^\dagger_a\mathcal{S}^{-1}=c_a
$
 on canonical Fermion annihilation and creation operators at the 
boundary\footnote{When defined
on a 1D lattice, one needs to include a factor $(-1)$ on the right hand side of this equation, when acting on Fermion operators
on one of the two sublattices, and a $(+1)$ sign, when acting on the other sublattice. 
Since this equation however refers only to one lattice site (at the boundary) we can
choose a $(+1)$ without loss of generality.}.
Fermion bilinear terms are again forbidden at the boundary by the $\U(1)\times Z_2^S$  symmetry  which protects that SPT order, so
that  the boundary Hamiltonian only contains charge-conserving  
interactions of fourth and higher order in Fermion operators that are invariant under the action of the chiral symmetry. The boundary Hamilonian
thus  roughly reads 
$ H=\sum V_{abcd}c^\dagger_{a}c^\dagger_{b}c_{c}c_{d}+\cdots$, or more precisely  
\eqs{\label{eq: SY} H=\sum_{a<b,c<d}&V_{abcd}[(c^\dagger_{a}c_{d}-\tfrac{1}{2}\delta_{ad})(c^\dagger_{b}c_{c}-\tfrac{1}{2}\delta_{bc})\\
&\phantom{V_{abcd}[}-(c^\dagger_{a}c_{c}-\tfrac{1}{2}\delta_{ac})(c^\dagger_{b}c_{d}-\tfrac{1}{2}\delta_{bd})]\\
&+h.c.+\cdots,} where the coefficients $V_{abcd}$ are complex numbers. In the above equation the  interaction terms are written in a way that makes their invariance under 
 chiral symmetry obvious. As in the BDI case, randomness in the  complex coefficients $V_{abcd}$  is induced by the randomness in the 1D bulk in symmetry class AIII (which, as discussed above, is here  non-interacting). Possible higher order random interactions are not written out explicitly in
\eqnref{eq: SY}. This model was first introduced by Sachdev and Ye\cite{Sachdev:1993lr}, and was revisited\cite{Sachdev:2015ao} recently in  view of its close analogy with the model in \eqnref{eq: Kitaev}, considered by Kitaev. We find that  the  level statistics of the Hamiltonian in \eqnref{eq: SY}  exhibits a  four-fold periodicity in $N_c$, matching the $\dsZ_4$ global anomaly on the boundary characteristic of class AIII with interactions. Let us explain our findings. Due to the $\U(1)$ symmetry, the Hamiltonian  can be  block-diagonalized in each $\U(1)$ charge sector, where the charge operator (with eigenvalue $q$)  reads  
\eq{\label{eq: Q} Q=\sum_{a=1}^{N_c}(c^\dagger_ac_a-1/2).}
 Therefore, the level statistics must be collected in each charge sector separately.  It turns out that there is an interplay between  the charge quantum number $q$
(= eigenvalue of $Q$)   and  the level statistics, as can be seen from our results  shown in \tabref{tab: AIII CII}(a): First, observe that there is an even-odd effect for   the charge $q$,  depending  on $N_c$: $q$ takes integer values if $N_c$ is even, and half-integer values if $N_c$ is odd (due to  charge fractionalization occuring at  the boundary of the  1D SPT). So the charge neutral sector ($q=0$) exists only for even $N_c$.  Second, in the $q=0$ sectors, the level statistics is that of GOE for $N_c (\mod 4)=0$ and of GSE for $N_c (\mod 4)=2$.  
(We can think\footnote{One may think of expessing the complex Fermions $c_a$ and $c^\dagger_a$ in terms of their Majorana ``real''-
and ``imaginary''- parts, which are those appearing in the BDI Hamiltonian. In this way, we can define the action of a $\U(1)$
on the BDI system,  we have $N_\chi=2 N_c$, and project that latter onto each $\U(1)$ charge sector.}
of the statistics as being inherited from the $N_\chi (\mod8)=0,4$ cases of class  
BDI 
via the correspondence  $N_\chi=2N_c$ -
compare Table \ref{tab: BDI}.).
Furthermore, in any $q\neq0$ sector,  
the chiral symmetry  operation $\mathcal{S}$ 
connects the  $q$ and $-q$  sectors which turns out to result in GUE level statistics  (see the discussion below).

\begin{table}[htbp]
\caption{Level statistics on the thermalized boundary of AIII and CII class MBL-SPT states. $q$ is the $\U(1)$ charge quantum number. The Fermion flavor number must be sufficiently large for the result to be universal.}  
\begin{center}
\begin{minipage}[t]{0.5\linewidth}
\centering  
\vspace{-8pt}
(a) AIII class\\
\vspace{4pt}
\begin{tabular}{c|cccc}
$N_c(\mod4)$ & 0 & 1 & 2 & 3\\
\hline  
$q=0$ & GOE & & GSE & \\
$q=\pm1/2$ &  & GUE & & GUE \\
$q\neq0$ & GUE & GUE & GUE & GUE  
\end{tabular}
\end{minipage}\hfill
\begin{minipage}[t]{0.4\linewidth}
\centering  
\vspace{-8pt}
(b) CII class\\
\vspace{4pt}
\begin{tabular}{c|cccc}
$N_f(\mod2)$ & 0 & 1 \\
\hline  
$q\in$\,even & GOE & GSE \\
$q\in$\,odd & GSE & GOE  
\end{tabular}
\end{minipage}
\end{center}
\label{tab: AIII CII}
\end{table}

The interplay of level statistics and symmetries can be understood from the analysis of  the projective representations of the 
chiral symmetry on the boundary. In general, the anti-unitary 
operator implementing the chiral symmetry on the many-body Fock space
can always we written as the  complex conjugation operator $\mathcal{K}$ followed by a unitary operator $\mathcal{U}$ on the Fock space,
i.e. $\mathcal{S}= \mathcal{U}   \mathcal{K}$.
The unitary operator $\mathcal{U}$ can be found by considering
its action 
on the boundary Fermions as follows. Let us first represent the Fermion operators $c_a$ as ``qubit operators'' $c_a=(\prod_{b<a}\sigma_b^z)(\sigma_a^x+\ii\sigma_a^y)/2$  using a Jordan-Wigner type  transformation. In this representation 
(i.e. in this basis of Fock space), $c_a$  and $c_a^\dagger$ are hence both 
 represented by  real matrices (using the standard convention for Pauli matrices). 
Both are therefore invariant under complex conjugation, $\mathcal{K}c_a\mathcal{K}^{-1}=c_a$,
and
$\mathcal{K}c^\dagger_a\mathcal{K}^{-1}=c^\dagger_a$,
 in this representation (i.e. in this basis of the Fermion Fock space). Therefore, to implement the chiral transformation on the Fermion operators, $\mathcal{S}c_a\mathcal{S}^{-1}=\mathcal{U}c_a \mathcal{U}^\dagger=c^\dagger_a$ 
and similarly
$\mathcal{U}c^\dagger_a \mathcal{U}^\dagger=c_a$,
one only needs to set  
\eq{\label{eq: T AIII} \mathcal{U}
\equiv
e^{i\pi N_c Q}
\prod_{a=1}^{N_c}\xi_a,
 \ \ {\rm where} \ \  \xi_a=\ii(c^\dagger_a-c_a)} 
are Majorana Fermion operators satisfying $\{\xi_a,\xi_b\}=2\delta_{ab}$, 
$\xi^\dagger_a=\xi_a$, as well as $\mathcal{K} \xi_a \mathcal{K}^{-1}
= -\xi_a$.
Here $Q$ is the $\U(1)$ charge operator defined in \eqnref{eq: Q}, which satisfies
$\mathcal{S} Q \mathcal{S}^{-1} = - Q$. Using these algebraic relations, it is straightforward to verify that 
\eq{ \label{eq:ChiralSymmetrySquaredFockSpace}\mathcal{S}^2=
\mathcal{U} \mathcal{U}^*= \bigg\{  
\begin{array}{ll}
+\mathbf{1} & \text{if }N_c\mod 4=0,1,\\
-\mathbf{1} & \text{if }N_c\mod 4=2,3,  
\end{array}}

\noindent
where $\mathcal{U}^* = \mathcal{K}\mathcal{U}\mathcal{K}^{-1}$.
Note that  since this result for $\mathcal{S}^2$ is invariant under a change of basis of the many-body Fock space,
it holds true in any such basis (even though it was initially derived in a representation in which both $c_a$ and $c^\dagger_a$
are real).\footnote{A general discussion of the relationship between the action of the square of anti-unitary operators such as chiral symmetry
or time-reversal symmetry on the many-body Fock space, and the action the same symmetry operations  on the single-particle Hilbert space, is provided in the Appendix. 
Let us briefly summarize the results for the chiral symmetry operation. The
action of these operators on the single-particle Hilbert space is determined by its action on the canonical Fermion operators.
For the chiral symmetry operator we can always make the choice that $\mathcal{S}^2 c_a \mathcal{S}^{-2}=c_a$,
$\mathcal{S}^2 c^\dagger_a \mathcal{S}^{-2}=c^\dagger_a$.  I.e., at the single-particle level the chiral symmetry operation
can always chosen to square to the identity (by a choice of phase).  It is shown in the Appendix that the square of the anti-unitary
operator $\mathcal{S}$ representing the chiral symmetry operation on the many-body Fermion Fock space, can nevertheless
have two possibilities, $\mathcal{S}^2 = \pm \mathbf{1}$, depending on the system. The result presented in
\eqnref{eq:ChiralSymmetrySquaredFockSpace}
is a particular example of the phenomenon discussed in the Appendix.}
Chiral symmetry
leaves the charge neutral sector ($q=0$) of the Hamiltonian invariant and is thus a symmetry of the Hamiltonian in this sector.
Specifically, in this sector, chiral symmetry 
 of  the Hamiltonian $H_{q=0}$  amounts to 
\eq{\label{eq: THT} \mathcal{S}H_{q=0}\mathcal{S}^{-1}=\mathcal{U}  H_{q=0}^* \mathcal{U}^{-1}=H_{q=0}.}
When  $N_c (\mod4) =0$ we have $\mathcal{U}\mathcal{U}^*=+\mathbf{1}$, and so one can choose 
a basis of the many-body Fock space in which
$\mathcal{U}=\mathbf{1}$.
 Then \eqnref{eq: THT} implies that $H_{q=0}\in\dsR$ is a real symmetric matrix, which should exhibit GOE level statistics in the ETH phase.
When
 $N_c (\mod4)=2$ we have
$\mathcal{U} \mathcal{U}^*=-\mathbf{1}$, and so one can choose
a basis of the many-body Fock space in which
$\mathbf{U}=\mat{0 &+\mathbf{1}\\ -\mathbf{1}&0}$.
Then \eqnref{eq: THT} implies that $H_{q=0}\in\dsH$ is a quaternion Hermitian matrix, which should consequently
exhibit GSE level statistics in the ETH phase.
Since, as mentioned above, $N_c$ must be even when $q=0$ this exhausts all possibilities for the $q=0$ sector.
However for $q\neq0$, 
the chiral symmetry transformation $\mathcal{S}$ connects the two
charge sectors $\pm q$.
In  block-matrix form, we have  
\eq{ Q=\mat{-q & 0\\0&+q}, \ \mathcal{U}=\mat{0& \mathbf{1}\\ \eta_{\mathcal{S}}\mathbf{1}&0}, \  H=\mat{H_{-q} & 0\\ 0& H_{+q}},}
 where $\eta_{\mathcal{S}}=\pm1$ depends on the projective representation $\mathcal{S}^2=\eta_{\mathcal{S}}$. 
But no matter what the value of  $\eta_{\mathcal{S}}$, \eqnref{eq: THT} only establishes a connection between 
$H_{+ q}$
and
$H_{-q}$, i.e. $H_{q}^*=H_{-q}$, which imposes no further restriction on $H_{q}$ itself. 
So for $q\neq 0$, $H_{q}\in\dsC$ is a complex Hermitian matrix, which should exhibit GUE level statistics in the ETH phase.\footnote{The above approach by analyzing the projective representations of the time-reversal symmetry can be applied to the BDI case as well. For more details, see the Appendixl \ref{sec: BDI}.}  - These predictions are confirmed by numerical studies of these spectra, and  displayed in \tabref{tab: AIII CII}(a). 

\vskip .1cm 

{\it Symmetry Class CII.}
 Now we turn to the MBL-SPT states in  symmetry class CII, which are protected by  
$(\U(1) \rtimes Z_2^{\mathcal{C}})\times Z_2^{\mathcal{S}}$ symmetry.
The $\dsZ$ classification of the  non-interacting 1D  SPT phases in this class reduces to  $\dsZ_2$ in the presence of
 interactions\footnote{This class of Fermionic SPT states  turns out to be also related to bosonic SPT states (Haldane chains)\cite{Senthil:2014qy,You:2014pt,You:2015bx} which have the same $\dsZ_2$ classification.}. The symmetry action on the boundary is 
understood most easily  if we embed the $\U(1)\rtimes Z_2^{\mathcal{C}}$ subgroup into the $\SU(2)$ group (although the $\SU(2)$ symmetry is not 
necessary\footnote{For $N_f=1$ (the ``root state'' of the CII class), 
there is an accidental $\SU(2)$ symmetry that emerges from $\U(1)\rtimes\dsZ_2^{\mathcal{C}}$. 
However for generic $N_f$ (with the large $N_f$ limit in mind), no $\SU(2)$ symmetry is required for the CII class. The introduction of $\SU(2)$ group here is merely a trick to help with the presentation.} 
to protect this SPT phase).
Therefore we
consider the boundary degrees of freedom to be spin-1/2 Fermions $f_a=(f_{a\uparrow},f_{a\downarrow})^T$,
where  $a=1,2,\cdots,N_f$. The $\SU(2)$ generators are defined as  
\eq{\label{eq: S}
\vect{\vec S}=\frac{1}{2}\sum_{a=1}^{N_f}f^\dagger_a\vect{\vec \sigma}f_a,}
where $\vect{\sigma}=(\sigma_x,\sigma_y,\sigma_z)$ are Pauli matrices. The $\U(1)$ symmetry in this representation of symmetry class  CII  corresponds to conservation of $S_z$, with the $\U(1)$ charge operator  
\eq{\label{eq:DefQSz} Q=2S_z=\sum_{a=1}^{N_f}
\sum_{\sigma = 1, 2}
 (-1)^\sigma f_{a\sigma}^\dagger f_{a\sigma}.} ``Charge''-conjugation corresponds to spin-rotation by angle $\pi$  about the $S_y$-axis  
\eq{\mathcal{C}=e^{\ii\pi S_y},}
so  that $\mathcal{C} f_{a\uparrow}\mathcal{C}^\dagger=f_{a\downarrow}$, 
$\mathcal{C} f_{a\downarrow} \mathcal{C}^\dagger= -f_{a\uparrow}$ and 
$\mathcal{C} Q \mathcal{C}^\dagger=-Q$ 
(which makes $\mathcal{C}$ consistent with its physical meaning of charge conjugation). The 
chiral  symmetry acts as $\mathcal{S}f_{a\sigma}\mathcal{S}^{-1}=f^\dagger_{a\sigma}$, which also flips the spin $\mathcal{S}\vect{\vec S}\mathcal{S}^{-1}=-\vect{\vec S}$, and in particular the ``charge'' $Q=2S_z$.
To implement the
chiral symmetry  operation, 
we write 
$\mathcal{S} = \mathcal{U} \mathcal{K}$ where $\mathcal{K}$
denotes complex conjugation and $\mathcal{U}$ a unitary operator in the many-body Fermion
Fock space.
In complete analogy with the AIII case discussed  above,
one first chooses again a real representation of the canonical Fermion operator $f_{a\sigma}$ and $f^\dagger_{a\sigma}$
using a Jordan-Wigner type transformation and ``qubit operators'', as was done  in the paragraph above
 \eqnref{eq: T AIII}, so that
$\mathcal{K} f_{a\sigma} \mathcal{K}^{-1}= f_{a\sigma}$, and 
$\mathcal{K} f^\dagger_{a\sigma} \mathcal{K}^{-1}= f^\dagger_{a\sigma}$.
As before, one immediately verifies that the action of the chiral symmetry transformation $\mathcal{S}$  on the Fermion operators is
reproduced by setting
\eq{\mathcal{U}=\prod_{a=1}^{N_f}\prod_{\sigma=\uparrow,\downarrow} \xi_{a\sigma}, \ \ {\rm where} \ \ 
\xi_{a\sigma}\equiv
\ii(f_{a\sigma}^\dagger-f_{a\sigma}),}
where again $\mathcal{K}\xi_{a\sigma}\mathcal{K}^{-1} = -\xi_{a\sigma}$.
One easily verifies
\begin{equation}
\label{LabelEqSquareOfChiralSymmetry}
\mathcal{S}^2 = \mathcal{U} \mathcal{U}^* = (-1)^{N_f} \ \ {\rm and} \ \ \mathcal{C}^2=(-1)^Q,
\end{equation}
where the 2nd equation follows directly from the action of $\mathcal{C}$ on the Fermion operators.
One also immediately verifies
the 
algebraic relations $\mathcal{S}e^{\ii\theta Q}=e^{\ii\theta Q}\mathcal{S}$
and  $\mathcal{S}\mathcal{C}=\mathcal{C}\mathcal{S}$ (such that $Z_2^{\mathcal{S}}$ commutes with
 $\U(1)\rtimes Z^{\mathcal{C}}$).
As in the case of AIII, note that since these relations, in particular
\eqnref{LabelEqSquareOfChiralSymmetry},
are  invariant under a change of basis of the many-body Fock space,
they hold true in any such basis (even though it was initially derived in a representation in which both $f_a$ and $f^\dagger_a$
are real).

The boundary Hamiltonian contains all $(\U(1)\rtimes Z_2^\mathcal{C})\times Z_2^\mathcal{S}$ symmetric random interactions. 
One may think of generating such   a Hamiltonian
from the Hamiltonian in symmetry class BDI appearing in \eqnref{eq: H BDI} containing an even number $N_\chi = 4 N_f$ of Majorana
Fermion species paired up so as to define the action of an $\U(1)$ symmetry, by projecting the latter
onto each $\U(1)$ charge sector and  by then  symmetrizing with respect to the $Z_2^\mathcal{C}$ group. (I.e., here we think
of expressing the complex Fermions $f_{a\sigma}$ and $f^\dagger_{a\sigma}$ in terms of their real and imaginary
parts Majorana Fermions, which are those appearing in the corresponding BDI Hamiltonian.
Recall the presence of the extra spin index, $N_c = 2 N_f$, when comparing to class AIII.)
We collect the level statistics numerically  in each charge sector labeled by the 
eigenvalue $q$ of the operator $Q$. It turns out that the GOE and GSE level statistics appear alternatively with respect to the parity of both the topological number $N_f$ and the charge quantum number $q$, as summarized in \tabref{tab: AIII CII}(b).   

Again, these numerically obtained results for the level statistics can be understood by analyzing the 
nature of the representations of the $(\U(1) \rtimes Z_2^{\mathcal{C}})\times Z_2^{\mathcal{S}}$
symmetry (which protects the SPT order).
In the charge neutral ($q=0$) sector,  charge conjugation $\mathcal{C}$ is effectively an identity operator. So the analysis is the same as the AIII case, which explains the GOE (or GSE) level statistics at $N_f (\mod2)=0$ (or $1$). For $q\neq 0$, opposite charge sectors $\pm q$ must 
again be put together for consideration since they are connected by the action of $\mathcal{C}$ and $\mathcal{S}$. In the block-diagonal basis of $Q$, we have  
\eq{Q=\mat{-q & 0\\0&+q}, \  \mathcal{C}=\mat{0 & \mathbf{1}\\ \eta_\mathcal{C}\mathbf{1} &0}, \  H=\mat{H_{-q} & 0\\ 0& H_{+q}}.}
The form of $\mathcal{C}$ is 
determined  by the relation $\mathcal{C}Q=-Q\mathcal{C}$, and $\eta_\mathcal{C}=\mathcal{C}^2=(-1)^q.$ 
To respect the $Z_2^\mathcal{C}$ symmetry, we require the Hamiltonian to satisfy $\mathcal{C}H=H\mathcal{C}$, which implies $H_{+q}=H_{-q}$. In the present  basis (block-diagonal in $Q$), the relations $\mathcal{S}e^{\ii\theta Q}=e^{\ii\theta Q}\mathcal{S}$ and $\mathcal{S}\mathcal{C}=\mathcal{C}\mathcal{S}$  translate into  $\mathcal{U} Q=-Q\mathcal{U}$ and $\mathcal{U} \mathcal{C}=\mathcal{C} \mathcal{U}$, 
so
that $\mathcal{U}$ must take the form of  
\eq{\label{eq: T CII} \mathcal{U}=\mat{0 & J \\ \eta_\mathcal{C} J &0},}
where $J$ is a real matrix to be determined.  
Upon substituting  \eqnref{eq: T CII} into $\mathcal{U} \mathcal{U}^*=\mathcal{S}^2=(-1)^{N_f}\equiv\eta_\mathcal{S}$, it is found that $J^2=\eta_\mathcal{C}\eta_\mathcal{S}=(-1)^{q+N_f}$.  In order to  respect the 
chiral  symmetry ($\mathcal{S}H\mathcal{S}^{-1}=H$), we must have  
\eq{\label{eq: JHJ} JH_{q}^*J^{-1}=H_{-q}=H_{q}.}
When $(q+N_f)$ is even (odd), $J^2=+\mathbf{1}$ ($-\mathbf{1}$), then \eqnref{eq: JHJ} implies
that  $H_q$ is
a real symmetric (quaternion Hermitian) matrix which leads to the GOE (GSE) level statistics. 
This result in combination
with the analysis in the $q=0$ sector, thus explains the numerical results displayed  in \tabref{tab: AIII CII}(b). One may extend the $(\U(1)\rtimes Z_2^\mathcal{C})\times Z_2^\mathcal{S}$ symmetry to an $\SU(2)\times Z_2^\mathcal{S}$ symmetry; the SPT classification and the level statistics remain the unchanged. With full $\SU(2)$ symmetry, the level statistics is to be  considered in each spin-$s$ sector, where the spin quantum number $s$  is determined by  $\vect{\vec S}^2=s(s+1)$. The even (odd) charge $q$  in \tabref{tab: AIII CII}(b) should then  be replaced by an integer (half-integer) spin $s$.   

An equivalent way of reading the above result arises  from  using the (many-body)  time-reversal operator, 
$\mathcal{T}=$ $\mathcal{S} \mathcal{C} =$ $\mathcal{C} \mathcal{S}$, whose square
becomes
 $\mathcal{T}^2 =$ $\mathcal{S}^2 \mathcal{C}^2=$ $(-1)^{N_f} \ (-1)^q$. Since the ``charge'' $q$ defines
the corresponding Fermion number parity operator $(-1)^F=  \ (-1)^q$,  the square of the many-body time reversal operator
is of the form
$\mathcal{T}^2 =$ $\gamma_\text{mb} \ (-1)^F$, where $\gamma_\text{mb} \equiv (-1)^{N_f}$ is a ``many-body'' phase that
may always appear when considering the time-reversal operator on the Fermionic (many-body) Fock space. For a general
discussion see \eqnref{LabelEqThetaSquareManyBody}, and the corresponding text in
the Appendix.\footnote{We end by commenting that we could have made a canonical transformation by making a particle-hole
transformation on the $\sigma = \downarrow$ Fermions (and not on the $\sigma = \uparrow$ Fermions):
$F_{a\uparrow}\equiv f_{a\uparrow}$, $F_{a\downarrow}\equiv f_{a\downarrow}$. Instead of \eqnref{eq:DefQSz}, we would
have obtained the usual expression $Q=$ $\sum_{a,\sigma} F^\dagger_{a\sigma}F_{a\sigma}$, while charge conjugation
and the action of the chiral symmetry
would also
have acquired the familiar forms
$\mathcal{C} F_{a\uparrow}\mathcal{C}^\dagger=F^\dagger_{a\downarrow}$, \  
and $\mathcal{C} F^\dagger_{a\downarrow} \mathcal{C}^\dagger= -F_{a\uparrow}$.
All the previous statements would
of course have been entirely identical after this canonical transformation.}

In conclusion, we have investigated the many-body level statistics of the SYK model for the three symmetry classes BDI, AIII and CII whose SPT phases in 1D are $\dsZ$ classified in the absence of interactions. The level statistics varies among the three different Wigner-Dyson random matrix ensembles periodically with the Fermion flavor number, which also corresponds to the topological index characterizing the interacting 1D SPT phases in these symmetry classes. There is an interesting interplay between level statistics and  symmetry quantum numbers, as summarized in \tabref{tab: BDI} and \tabref{tab: AIII CII}. The patterns of level statistics can be understood from the global quantum anomalies which are known to characterize the 1D bulk SPT phases, by considering the SYK models as effective theories for the thermalized boundaries of 1D Fermionic MBL-SPT states.  

As our work was completed, we became aware of 
results by 
Fu and Sachdev\cite{Sachdev:ly} who also discover the $\dsZ_4$ periodicity of the SY model (symmetry class AIII)
 in the level degeneracy. We are grateful to Wenbo Fu for sharing his unpublished results with us at that point. We also acknowledge helpful discussions with Andrew Potter, Tarun Grover, Xiao-Liang Qi and Marcos Rigol. YZY and CX are supported by the David and Lucile Packard foundation and NSF Grant No. DMR-1151208. AWWL is supported by NSF Grant No. DMR-1309667. 

\bibliography{levstatNew3}

\begin{thebibliography}{75}
\expandafter\ifx\csname natexlab\endcsname\relax\def\natexlab#1{#1}\fi
\expandafter\ifx\csname bibnamefont\endcsname\relax
  \def\bibnamefont#1{#1}\fi
\expandafter\ifx\csname bibfnamefont\endcsname\relax
  \def\bibfnamefont#1{#1}\fi
\expandafter\ifx\csname citenamefont\endcsname\relax
  \def\citenamefont#1{#1}\fi
\expandafter\ifx\csname url\endcsname\relax
  \def\url#1{\texttt{#1}}\fi
\expandafter\ifx\csname urlprefix\endcsname\relax\def\urlprefix{URL }\fi
\providecommand{\bibinfo}[2]{#2}
\providecommand{\eprint}[2][]{\url{#2}}

\bibitem[{\citenamefont{{Sachdev} and {Ye}}(1993)}]{Sachdev:1993lr}
\bibinfo{author}{\bibfnamefont{S.}~\bibnamefont{{Sachdev}}} \bibnamefont{and}
  \bibinfo{author}{\bibfnamefont{J.}~\bibnamefont{{Ye}}},
  \bibinfo{journal}{Physical Review Letters} \textbf{\bibinfo{volume}{70}},
  \bibinfo{pages}{3339} (\bibinfo{year}{1993}), \eprint{cond-mat/9212030}.

\bibitem[{\citenamefont{{Sachdev}}(2015)}]{Sachdev:2015ao}
\bibinfo{author}{\bibfnamefont{S.}~\bibnamefont{{Sachdev}}},
  \bibinfo{journal}{Physical Review X} \textbf{\bibinfo{volume}{5}},
  \bibinfo{eid}{041025} (\bibinfo{year}{2015}), \eprint{1506.05111}.

\bibitem[{\citenamefont{{Kitaev}}(2015)}]{Kitaev:2015uj}
\bibinfo{author}{\bibfnamefont{A.}~\bibnamefont{{Kitaev}}}
  (\bibinfo{year}{2015}), \bibinfo{note}{talk at KITP Program: Entanglement in
  Strongly-Correlated Quantum Matter},
  \urlprefix\url{http://online.kitp.ucsb.edu/online/entangled15/kitaev/}.

\bibitem[{\citenamefont{Gu and Wen}(2009)}]{Wen:2009qv}
\bibinfo{author}{\bibfnamefont{Z.-C.} \bibnamefont{Gu}} \bibnamefont{and}
  \bibinfo{author}{\bibfnamefont{X.-G.} \bibnamefont{Wen}},
  \bibinfo{journal}{Phys. Rev. B} \textbf{\bibinfo{volume}{80}},
  \bibinfo{pages}{155131} (\bibinfo{year}{2009}), \eprint{0903.1069}.

\bibitem[{\citenamefont{Chen et~al.}(2011)\citenamefont{Chen, Gu, and
  Wen}}]{Wen:2011sf}
\bibinfo{author}{\bibfnamefont{X.}~\bibnamefont{Chen}},
  \bibinfo{author}{\bibfnamefont{Z.-C.} \bibnamefont{Gu}}, \bibnamefont{and}
  \bibinfo{author}{\bibfnamefont{X.-G.} \bibnamefont{Wen}},
  \bibinfo{journal}{Phys. Rev. B} \textbf{\bibinfo{volume}{83}},
  \bibinfo{pages}{035107} (\bibinfo{year}{2011}), \eprint{1008.3745}.

\bibitem[{\citenamefont{{Fidkowski} and {Kitaev}}(2011)}]{Fidkowski:2011pa}
\bibinfo{author}{\bibfnamefont{L.}~\bibnamefont{{Fidkowski}}} \bibnamefont{and}
  \bibinfo{author}{\bibfnamefont{A.}~\bibnamefont{{Kitaev}}},
  \bibinfo{journal}{\prb} \textbf{\bibinfo{volume}{83}}, \bibinfo{eid}{075103}
  (\bibinfo{year}{2011}), \eprint{1008.4138}.

\bibitem[{\citenamefont{{Haldane}}(1983)}]{Haldane:1983fu}
\bibinfo{author}{\bibfnamefont{F.~D.~M.} \bibnamefont{{Haldane}}},
  \bibinfo{journal}{Physics Letters A} \textbf{\bibinfo{volume}{93}},
  \bibinfo{pages}{464} (\bibinfo{year}{1983}).

\bibitem[{\citenamefont{Haldane}(1983)}]{Haldane:1983vl}
\bibinfo{author}{\bibfnamefont{F.~D.~M.} \bibnamefont{Haldane}},
  \bibinfo{journal}{Phys. Rev. Lett.} \textbf{\bibinfo{volume}{50}},
  \bibinfo{pages}{1153} (\bibinfo{year}{1983}).

\bibitem[{\citenamefont{Affleck et~al.}(1987)\citenamefont{Affleck, Kennedy,
  Lieb, and Tasaki}}]{AKLT:1987dp}
\bibinfo{author}{\bibfnamefont{I.}~\bibnamefont{Affleck}},
  \bibinfo{author}{\bibfnamefont{T.}~\bibnamefont{Kennedy}},
  \bibinfo{author}{\bibfnamefont{E.~H.} \bibnamefont{Lieb}}, \bibnamefont{and}
  \bibinfo{author}{\bibfnamefont{H.}~\bibnamefont{Tasaki}},
  \bibinfo{journal}{Phys. Rev. Lett.} \textbf{\bibinfo{volume}{59}},
  \bibinfo{pages}{799} (\bibinfo{year}{1987}).

\bibitem[{\citenamefont{{Kitaev}}(2001)}]{Kitaev:2001un}
\bibinfo{author}{\bibfnamefont{A.~Y.} \bibnamefont{{Kitaev}}},
  \bibinfo{journal}{Physics Uspekhi} \textbf{\bibinfo{volume}{44}},
  \bibinfo{pages}{131} (\bibinfo{year}{2001}), \eprint{cond-mat/0010440}.

\bibitem[{\citenamefont{Read and Green}(2000)}]{Read:2000sc}
\bibinfo{author}{\bibfnamefont{N.}~\bibnamefont{Read}} \bibnamefont{and}
  \bibinfo{author}{\bibfnamefont{D.}~\bibnamefont{Green}},
  \bibinfo{journal}{Phys. Rev. B} \textbf{\bibinfo{volume}{61}},
  \bibinfo{pages}{10267} (\bibinfo{year}{2000}).

\bibitem[{\citenamefont{{Kane} and {Mele}}(2005)}]{Kane:2005lr}
\bibinfo{author}{\bibfnamefont{C.~L.} \bibnamefont{{Kane}}} \bibnamefont{and}
  \bibinfo{author}{\bibfnamefont{E.~J.} \bibnamefont{{Mele}}},
  \bibinfo{journal}{Physical Review Letters} \textbf{\bibinfo{volume}{95}},
  \bibinfo{eid}{146802} (\bibinfo{year}{2005}), \eprint{cond-mat/0506581}.

\bibitem[{\citenamefont{{Moore} and {Balents}}(2007)}]{Moore:2007ao}
\bibinfo{author}{\bibfnamefont{J.~E.} \bibnamefont{{Moore}}} \bibnamefont{and}
  \bibinfo{author}{\bibfnamefont{L.}~\bibnamefont{{Balents}}},
  \bibinfo{journal}{\prb} \textbf{\bibinfo{volume}{75}}, \bibinfo{eid}{121306}
  (\bibinfo{year}{2007}), \eprint{cond-mat/0607314}.

\bibitem[{\citenamefont{Hasan and Kane}(2010)}]{Hasan:2010rv}
\bibinfo{author}{\bibfnamefont{M.~Z.} \bibnamefont{Hasan}} \bibnamefont{and}
  \bibinfo{author}{\bibfnamefont{C.~L.} \bibnamefont{Kane}},
  \bibinfo{journal}{Rev. Mod. Phys.} \textbf{\bibinfo{volume}{82}},
  \bibinfo{pages}{3045} (\bibinfo{year}{2010}), \eprint{1002.3895}.

\bibitem[{\citenamefont{{Qi} and {Zhang}}(2011)}]{Qi:2011rv}
\bibinfo{author}{\bibfnamefont{X.-L.} \bibnamefont{{Qi}}} \bibnamefont{and}
  \bibinfo{author}{\bibfnamefont{S.-C.} \bibnamefont{{Zhang}}},
  \bibinfo{journal}{Reviews of Modern Physics} \textbf{\bibinfo{volume}{83}},
  \bibinfo{pages}{1057} (\bibinfo{year}{2011}), \eprint{1008.2026}.

\bibitem[{\citenamefont{{Schnyder} et~al.}(2008)\citenamefont{{Schnyder},
  {Ryu}, {Furusaki}, and {Ludwig}}}]{Ludwig:2008zy}
\bibinfo{author}{\bibfnamefont{A.~P.} \bibnamefont{{Schnyder}}},
  \bibinfo{author}{\bibfnamefont{S.}~\bibnamefont{{Ryu}}},
  \bibinfo{author}{\bibfnamefont{A.}~\bibnamefont{{Furusaki}}},
  \bibnamefont{and} \bibinfo{author}{\bibfnamefont{A.~W.~W.}
  \bibnamefont{{Ludwig}}}, \bibinfo{journal}{\prb}
  \textbf{\bibinfo{volume}{78}}, \bibinfo{eid}{195125} (\bibinfo{year}{2008}),
  \eprint{0803.2786}.

\bibitem[{\citenamefont{{Ryu} et~al.}(2010)\citenamefont{{Ryu}, {Schnyder},
  {Furusaki}, and {Ludwig}}}]{Ludwig:2010xt}
\bibinfo{author}{\bibfnamefont{S.}~\bibnamefont{{Ryu}}},
  \bibinfo{author}{\bibfnamefont{A.~P.} \bibnamefont{{Schnyder}}},
  \bibinfo{author}{\bibfnamefont{A.}~\bibnamefont{{Furusaki}}},
  \bibnamefont{and} \bibinfo{author}{\bibfnamefont{A.~W.~W.}
  \bibnamefont{{Ludwig}}}, \bibinfo{journal}{New Journal of Physics}
  \textbf{\bibinfo{volume}{12}}, \bibinfo{eid}{065010} (\bibinfo{year}{2010}),
  \eprint{0912.2157}.

\bibitem[{\citenamefont{{Kitaev}}(2009)}]{Kitaev:2009sd}
\bibinfo{author}{\bibfnamefont{A.}~\bibnamefont{{Kitaev}}}, in
  \emph{\bibinfo{booktitle}{American Institute of Physics Conference Series}},
  edited by \bibinfo{editor}{\bibfnamefont{V.}~\bibnamefont{{Lebedev}}}
  \bibnamefont{and}
  \bibinfo{editor}{\bibfnamefont{M.}~\bibnamefont{{Feigel'Man}}}
  (\bibinfo{year}{2009}), vol. \bibinfo{volume}{1134} of
  \emph{\bibinfo{series}{American Institute of Physics Conference Series}}, pp.
  \bibinfo{pages}{22--30}, \eprint{0901.2686}.

\bibitem[{\citenamefont{{Ludwig}}(2015{\natexlab{a}})}]{Ludwig:2015la}
\bibinfo{author}{\bibfnamefont{A.~W.~W.} \bibnamefont{{Ludwig}}},
  \bibinfo{journal}{Physica Scripta} \textbf{\bibinfo{volume}{2016}},
  \bibinfo{pages}{014001} (\bibinfo{year}{2015}{\natexlab{a}}).

\bibitem[{\citenamefont{Deutsch}(1991)}]{Deutsch:1991ik}
\bibinfo{author}{\bibfnamefont{J.~M.} \bibnamefont{Deutsch}},
  \bibinfo{journal}{Phys. Rev. A} \textbf{\bibinfo{volume}{43}},
  \bibinfo{pages}{2046} (\bibinfo{year}{1991}).

\bibitem[{\citenamefont{Srednicki}(1994)}]{Srednicki:1994ns}
\bibinfo{author}{\bibfnamefont{M.}~\bibnamefont{Srednicki}},
  \bibinfo{journal}{Phys. Rev. E} \textbf{\bibinfo{volume}{50}},
  \bibinfo{pages}{888} (\bibinfo{year}{1994}), \eprint{cond-mat/9403051}.

\bibitem[{\citenamefont{{Rigol} et~al.}(2008)\citenamefont{{Rigol}, {Dunjko},
  and {Olshanii}}}]{Rigol:2008qr}
\bibinfo{author}{\bibfnamefont{M.}~\bibnamefont{{Rigol}}},
  \bibinfo{author}{\bibfnamefont{V.}~\bibnamefont{{Dunjko}}}, \bibnamefont{and}
  \bibinfo{author}{\bibfnamefont{M.}~\bibnamefont{{Olshanii}}},
  \bibinfo{journal}{\nat} \textbf{\bibinfo{volume}{452}}, \bibinfo{pages}{854}
  (\bibinfo{year}{2008}), \eprint{0708.1324}.

\bibitem[{\citenamefont{{Santos} et~al.}(2012)\citenamefont{{Santos},
  {Polkovnikov}, and {Rigol}}}]{Rigol:2012wd}
\bibinfo{author}{\bibfnamefont{L.~F.} \bibnamefont{{Santos}}},
  \bibinfo{author}{\bibfnamefont{A.}~\bibnamefont{{Polkovnikov}}},
  \bibnamefont{and} \bibinfo{author}{\bibfnamefont{M.}~\bibnamefont{{Rigol}}},
  \bibinfo{journal}{\pre} \textbf{\bibinfo{volume}{86}}, \bibinfo{eid}{010102}
  (\bibinfo{year}{2012}), \eprint{1202.4764}.

\bibitem[{\citenamefont{{Basko} et~al.}(2006)\citenamefont{{Basko}, {Aleiner},
  and {Altshuler}}}]{BAA:2006}
\bibinfo{author}{\bibfnamefont{D.~M.} \bibnamefont{{Basko}}},
  \bibinfo{author}{\bibfnamefont{I.~L.} \bibnamefont{{Aleiner}}},
  \bibnamefont{and} \bibinfo{author}{\bibfnamefont{B.~L.}
  \bibnamefont{{Altshuler}}}, \bibinfo{journal}{Annals of Physics}
  \textbf{\bibinfo{volume}{321}}, \bibinfo{pages}{1126} (\bibinfo{year}{2006}),
  \eprint{cond-mat/0506617}.

\bibitem[{\citenamefont{{Gornyi} et~al.}(2005)\citenamefont{{Gornyi}, {Mirlin},
  and {Polyakov}}}]{GMP:2005}
\bibinfo{author}{\bibfnamefont{I.~V.} \bibnamefont{{Gornyi}}},
  \bibinfo{author}{\bibfnamefont{A.~D.} \bibnamefont{{Mirlin}}},
  \bibnamefont{and} \bibinfo{author}{\bibfnamefont{D.~G.}
  \bibnamefont{{Polyakov}}}, \bibinfo{journal}{Physical Review Letters}
  \textbf{\bibinfo{volume}{95}}, \bibinfo{eid}{206603} (\bibinfo{year}{2005}),
  \eprint{cond-mat/0506411}.

\bibitem[{\citenamefont{{Oganesyan} and {Huse}}(2007)}]{Huse:2007is}
\bibinfo{author}{\bibfnamefont{V.}~\bibnamefont{{Oganesyan}}} \bibnamefont{and}
  \bibinfo{author}{\bibfnamefont{D.~A.} \bibnamefont{{Huse}}},
  \bibinfo{journal}{\prb} \textbf{\bibinfo{volume}{75}}, \bibinfo{eid}{155111}
  (\bibinfo{year}{2007}), \eprint{cond-mat/0610854}.

\bibitem[{\citenamefont{{{\v Z}nidari{\v c}} et~al.}(2008)\citenamefont{{{\v
  Z}nidari{\v c}}, {Prosen}, and {Prelov{\v s}ek}}}]{Znidaric:2008he}
\bibinfo{author}{\bibfnamefont{M.}~\bibnamefont{{{\v Z}nidari{\v c}}}},
  \bibinfo{author}{\bibfnamefont{T.}~\bibnamefont{{Prosen}}}, \bibnamefont{and}
  \bibinfo{author}{\bibfnamefont{P.}~\bibnamefont{{Prelov{\v s}ek}}},
  \bibinfo{journal}{\prb} \textbf{\bibinfo{volume}{77}}, \bibinfo{eid}{064426}
  (\bibinfo{year}{2008}), \eprint{0706.2539}.

\bibitem[{\citenamefont{{Imbrie}}(2014)}]{Imbrie:2014}
\bibinfo{author}{\bibfnamefont{J.~Z.} \bibnamefont{{Imbrie}}},
  \bibinfo{journal}{ArXiv e-prints}  (\bibinfo{year}{2014}),
  \eprint{1403.7837}.

\bibitem[{\citenamefont{{Nandkishore} and {Huse}}(2015)}]{Huse:2015rv}
\bibinfo{author}{\bibfnamefont{R.}~\bibnamefont{{Nandkishore}}}
  \bibnamefont{and} \bibinfo{author}{\bibfnamefont{D.~A.}
  \bibnamefont{{Huse}}}, \bibinfo{journal}{Annual Review of Condensed Matter
  Physics} \textbf{\bibinfo{volume}{6}}, \bibinfo{pages}{15}
  (\bibinfo{year}{2015}), \eprint{1404.0686}.

\bibitem[{\citenamefont{{Bauer} and {Nayak}}(2013)}]{Bauer:2013al}
\bibinfo{author}{\bibfnamefont{B.}~\bibnamefont{{Bauer}}} \bibnamefont{and}
  \bibinfo{author}{\bibfnamefont{C.}~\bibnamefont{{Nayak}}},
  \bibinfo{journal}{Journal of Statistical Mechanics: Theory and Experiment}
  \textbf{\bibinfo{volume}{9}}, \bibinfo{eid}{09005} (\bibinfo{year}{2013}),
  \eprint{1306.5753}.

\bibitem[{\citenamefont{{Huse} et~al.}(2013)\citenamefont{{Huse},
  {Nandkishore}, {Oganesyan}, {Pal}, and {Sondhi}}}]{Huse:2013ci}
\bibinfo{author}{\bibfnamefont{D.~A.} \bibnamefont{{Huse}}},
  \bibinfo{author}{\bibfnamefont{R.}~\bibnamefont{{Nandkishore}}},
  \bibinfo{author}{\bibfnamefont{V.}~\bibnamefont{{Oganesyan}}},
  \bibinfo{author}{\bibfnamefont{A.}~\bibnamefont{{Pal}}}, \bibnamefont{and}
  \bibinfo{author}{\bibfnamefont{S.~L.} \bibnamefont{{Sondhi}}},
  \bibinfo{journal}{\prb} \textbf{\bibinfo{volume}{88}}, \bibinfo{eid}{014206}
  (\bibinfo{year}{2013}), \eprint{1304.1158}.

\bibitem[{\citenamefont{{Bahri} et~al.}(2013)\citenamefont{{Bahri}, {Vosk},
  {Altman}, and {Vishwanath}}}]{Vishwanath:2013tp}
\bibinfo{author}{\bibfnamefont{Y.}~\bibnamefont{{Bahri}}},
  \bibinfo{author}{\bibfnamefont{R.}~\bibnamefont{{Vosk}}},
  \bibinfo{author}{\bibfnamefont{E.}~\bibnamefont{{Altman}}}, \bibnamefont{and}
  \bibinfo{author}{\bibfnamefont{A.}~\bibnamefont{{Vishwanath}}},
  \bibinfo{journal}{ArXiv e-prints}  (\bibinfo{year}{2013}),
  \eprint{1307.4092}.

\bibitem[{\citenamefont{{Pekker} et~al.}(2014)\citenamefont{{Pekker}, {Refael},
  {Altman}, {Demler}, and {Oganesyan}}}]{Altman:2014hg}
\bibinfo{author}{\bibfnamefont{D.}~\bibnamefont{{Pekker}}},
  \bibinfo{author}{\bibfnamefont{G.}~\bibnamefont{{Refael}}},
  \bibinfo{author}{\bibfnamefont{E.}~\bibnamefont{{Altman}}},
  \bibinfo{author}{\bibfnamefont{E.}~\bibnamefont{{Demler}}}, \bibnamefont{and}
  \bibinfo{author}{\bibfnamefont{V.}~\bibnamefont{{Oganesyan}}},
  \bibinfo{journal}{Physical Review X} \textbf{\bibinfo{volume}{4}},
  \bibinfo{eid}{011052} (\bibinfo{year}{2014}), \eprint{1307.3253}.

\bibitem[{\citenamefont{{Chandran} et~al.}(2014)\citenamefont{{Chandran},
  {Khemani}, {Laumann}, and {Sondhi}}}]{Chandran:2014fv}
\bibinfo{author}{\bibfnamefont{A.}~\bibnamefont{{Chandran}}},
  \bibinfo{author}{\bibfnamefont{V.}~\bibnamefont{{Khemani}}},
  \bibinfo{author}{\bibfnamefont{C.~R.} \bibnamefont{{Laumann}}},
  \bibnamefont{and} \bibinfo{author}{\bibfnamefont{S.~L.}
  \bibnamefont{{Sondhi}}}, \bibinfo{journal}{\prb}
  \textbf{\bibinfo{volume}{89}}, \bibinfo{eid}{144201} (\bibinfo{year}{2014}),
  \eprint{1310.1096}.

\bibitem[{\citenamefont{{Vosk} and {Altman}}(2014)}]{Altman:2014dq}
\bibinfo{author}{\bibfnamefont{R.}~\bibnamefont{{Vosk}}} \bibnamefont{and}
  \bibinfo{author}{\bibfnamefont{E.}~\bibnamefont{{Altman}}},
  \bibinfo{journal}{Physical Review Letters} \textbf{\bibinfo{volume}{112}},
  \bibinfo{eid}{217204} (\bibinfo{year}{2014}), \eprint{1307.3256}.

\bibitem[{\citenamefont{{Potter} and {Vishwanath}}(2015)}]{Vishwanath:2015po}
\bibinfo{author}{\bibfnamefont{A.~C.} \bibnamefont{{Potter}}} \bibnamefont{and}
  \bibinfo{author}{\bibfnamefont{A.}~\bibnamefont{{Vishwanath}}},
  \bibinfo{journal}{ArXiv e-prints}  (\bibinfo{year}{2015}),
  \eprint{1506.00592}.

\bibitem[{\citenamefont{{Slagle} et~al.}(2015)\citenamefont{{Slagle}, {Bi},
  {You}, and {Xu}}}]{Xu:2015gg}
\bibinfo{author}{\bibfnamefont{K.}~\bibnamefont{{Slagle}}},
  \bibinfo{author}{\bibfnamefont{Z.}~\bibnamefont{{Bi}}},
  \bibinfo{author}{\bibfnamefont{Y.-Z.} \bibnamefont{{You}}}, \bibnamefont{and}
  \bibinfo{author}{\bibfnamefont{C.}~\bibnamefont{{Xu}}},
  \bibinfo{journal}{ArXiv e-prints}  (\bibinfo{year}{2015}),
  \eprint{1505.05147}.

\bibitem[{\citenamefont{{Serbyn}
  et~al.}(2013{\natexlab{a}})\citenamefont{{Serbyn}, {Papi{\'c}}, and
  {Abanin}}}]{Abanin:2013ta}
\bibinfo{author}{\bibfnamefont{M.}~\bibnamefont{{Serbyn}}},
  \bibinfo{author}{\bibfnamefont{Z.}~\bibnamefont{{Papi{\'c}}}},
  \bibnamefont{and} \bibinfo{author}{\bibfnamefont{D.~A.}
  \bibnamefont{{Abanin}}}, \bibinfo{journal}{Physical Review Letters}
  \textbf{\bibinfo{volume}{110}}, \bibinfo{eid}{260601}
  (\bibinfo{year}{2013}{\natexlab{a}}), \eprint{1304.4605}.

\bibitem[{\citenamefont{{Serbyn}
  et~al.}(2013{\natexlab{b}})\citenamefont{{Serbyn}, {Papi{\'c}}, and
  {Abanin}}}]{Abanin:2013lc}
\bibinfo{author}{\bibfnamefont{M.}~\bibnamefont{{Serbyn}}},
  \bibinfo{author}{\bibfnamefont{Z.}~\bibnamefont{{Papi{\'c}}}},
  \bibnamefont{and} \bibinfo{author}{\bibfnamefont{D.~A.}
  \bibnamefont{{Abanin}}}, \bibinfo{journal}{Physical Review Letters}
  \textbf{\bibinfo{volume}{111}}, \bibinfo{eid}{127201}
  (\bibinfo{year}{2013}{\natexlab{b}}), \eprint{1305.5554}.

\bibitem[{\citenamefont{{Huse} et~al.}(2014)\citenamefont{{Huse},
  {Nandkishore}, and {Oganesyan}}}]{Huse:2014ec}
\bibinfo{author}{\bibfnamefont{D.~A.} \bibnamefont{{Huse}}},
  \bibinfo{author}{\bibfnamefont{R.}~\bibnamefont{{Nandkishore}}},
  \bibnamefont{and}
  \bibinfo{author}{\bibfnamefont{V.}~\bibnamefont{{Oganesyan}}},
  \bibinfo{journal}{\prb} \textbf{\bibinfo{volume}{90}}, \bibinfo{eid}{174202}
  (\bibinfo{year}{2014}), \eprint{1305.4915}.

\bibitem[{\citenamefont{{Ros} et~al.}(2015)\citenamefont{{Ros}, {M{\"u}ller},
  and {Scardicchio}}}]{Scardicchio:2015kl}
\bibinfo{author}{\bibfnamefont{V.}~\bibnamefont{{Ros}}},
  \bibinfo{author}{\bibfnamefont{M.}~\bibnamefont{{M{\"u}ller}}},
  \bibnamefont{and}
  \bibinfo{author}{\bibfnamefont{A.}~\bibnamefont{{Scardicchio}}},
  \bibinfo{journal}{Nuclear Physics B} \textbf{\bibinfo{volume}{891}},
  \bibinfo{pages}{420} (\bibinfo{year}{2015}), \eprint{1406.2175}.

\bibitem[{\citenamefont{{Chandran} et~al.}(2015)\citenamefont{{Chandran},
  {Kim}, {Vidal}, and {Abanin}}}]{Abanin:2015io}
\bibinfo{author}{\bibfnamefont{A.}~\bibnamefont{{Chandran}}},
  \bibinfo{author}{\bibfnamefont{I.~H.} \bibnamefont{{Kim}}},
  \bibinfo{author}{\bibfnamefont{G.}~\bibnamefont{{Vidal}}}, \bibnamefont{and}
  \bibinfo{author}{\bibfnamefont{D.~A.} \bibnamefont{{Abanin}}},
  \bibinfo{journal}{\prb} \textbf{\bibinfo{volume}{91}}, \bibinfo{eid}{085425}
  (\bibinfo{year}{2015}), \eprint{1407.8480}.

\bibitem[{\citenamefont{{Rademaker}}(2015)}]{Rademaker:2015ve}
\bibinfo{author}{\bibfnamefont{L.}~\bibnamefont{{Rademaker}}},
  \bibinfo{journal}{ArXiv e-prints}  (\bibinfo{year}{2015}),
  \eprint{1507.07276}.

\bibitem[{\citenamefont{{You} et~al.}(2015)\citenamefont{{You}, {Qi}, and
  {Xu}}}]{You:2015sb}
\bibinfo{author}{\bibfnamefont{Y.-Z.} \bibnamefont{{You}}},
  \bibinfo{author}{\bibfnamefont{X.-L.} \bibnamefont{{Qi}}}, \bibnamefont{and}
  \bibinfo{author}{\bibfnamefont{C.}~\bibnamefont{{Xu}}},
  \bibinfo{journal}{ArXiv e-prints}  (\bibinfo{year}{2015}),
  \eprint{1508.03635}.

\bibitem[{\citenamefont{{Ryu} et~al.}(2012)\citenamefont{{Ryu}, {Moore}, and
  {Ludwig}}}]{RyuMooreLudwig:2012PRB}
\bibinfo{author}{\bibfnamefont{S.}~\bibnamefont{{Ryu}}},
  \bibinfo{author}{\bibfnamefont{J.~M.} \bibnamefont{{Moore}}},
  \bibnamefont{and} \bibinfo{author}{\bibfnamefont{A.~W.~W.}
  \bibnamefont{{Ludwig}}}, \bibinfo{journal}{\prb}
  \textbf{\bibinfo{volume}{85}}, \bibinfo{eid}{045104} (\bibinfo{year}{2012}),
  \eprint{1010.0936}.

\bibitem[{\citenamefont{{Wen}}(2013)}]{Wen:2013qf}
\bibinfo{author}{\bibfnamefont{X.-G.} \bibnamefont{{Wen}}},
  \bibinfo{journal}{\prd} \textbf{\bibinfo{volume}{88}}, \bibinfo{eid}{045013}
  (\bibinfo{year}{2013}), \eprint{1303.1803}.

\bibitem[{\citenamefont{{Ryu}}(2015)}]{RyuNobelSymp2015}
\bibinfo{author}{\bibfnamefont{S.}~\bibnamefont{{Ryu}}},
  \bibinfo{journal}{Physica Scripta} \bibinfo{eid}{014009}
  (\bibinfo{year}{2015}).

\bibitem[{\citenamefont{{Ludwig}}(2015{\natexlab{b}})}]{LudwigNobelSymp2015}
\bibinfo{author}{\bibfnamefont{A.~W.~W.} \bibnamefont{{Ludwig}}},
  \bibinfo{journal}{arXiv:1512.08882}  (\bibinfo{year}{2015}{\natexlab{b}}).

\bibitem[{\citenamefont{{Witten}}(2015)}]{WittenTopPhases2015}
\bibinfo{author}{\bibfnamefont{E.}~\bibnamefont{{Witten}}},
  \bibinfo{journal}{arXiv:1508.04715}  (\bibinfo{year}{2015}).

\bibitem[{\citenamefont{Altland and Zirnbauer}(1997)}]{AZclass}
\bibinfo{author}{\bibfnamefont{A.}~\bibnamefont{Altland}} \bibnamefont{and}
  \bibinfo{author}{\bibfnamefont{M.~R.} \bibnamefont{Zirnbauer}},
  \bibinfo{journal}{Phys. Rev. B} \textbf{\bibinfo{volume}{55}},
  \bibinfo{pages}{1142} (\bibinfo{year}{1997}).

\bibitem[{\citenamefont{{Heinzner} et~al.}(2005)\citenamefont{{Heinzner},
  {Huckleberry}, and {Zirnbauer}}}]{HeinznerCMP2005}
\bibinfo{author}{\bibfnamefont{P.}~\bibnamefont{{Heinzner}}},
  \bibinfo{author}{\bibfnamefont{A.}~\bibnamefont{{Huckleberry}}},
  \bibnamefont{and} \bibinfo{author}{\bibfnamefont{M.~R.}
  \bibnamefont{{Zirnbauer}}}, \bibinfo{journal}{Commun. Math. Phys.}
  \textbf{\bibinfo{volume}{257}}, \bibinfo{pages}{725} (\bibinfo{year}{2005}).

\bibitem[{\citenamefont{{Vosk} et~al.}(2014)\citenamefont{{Vosk}, {Huse}, and
  {Altman}}}]{Huse:2014cs}
\bibinfo{author}{\bibfnamefont{R.}~\bibnamefont{{Vosk}}},
  \bibinfo{author}{\bibfnamefont{D.~A.} \bibnamefont{{Huse}}},
  \bibnamefont{and} \bibinfo{author}{\bibfnamefont{E.}~\bibnamefont{{Altman}}},
  \bibinfo{journal}{ArXiv e-prints}  (\bibinfo{year}{2014}),
  \eprint{1412.3117}.

\bibitem[{\citenamefont{{Potter} et~al.}(2015)\citenamefont{{Potter},
  {Vasseur}, and {Parameswaran}}}]{Potter:2015kl}
\bibinfo{author}{\bibfnamefont{A.~C.} \bibnamefont{{Potter}}},
  \bibinfo{author}{\bibfnamefont{R.}~\bibnamefont{{Vasseur}}},
  \bibnamefont{and} \bibinfo{author}{\bibfnamefont{S.~A.}
  \bibnamefont{{Parameswaran}}}, \bibinfo{journal}{ArXiv e-prints}
  (\bibinfo{year}{2015}), \eprint{1501.03501}.

\bibitem[{\citenamefont{{Fidkowski} and {Kitaev}}(2010)}]{Fidkowski:2010iv}
\bibinfo{author}{\bibfnamefont{L.}~\bibnamefont{{Fidkowski}}} \bibnamefont{and}
  \bibinfo{author}{\bibfnamefont{A.}~\bibnamefont{{Kitaev}}},
  \bibinfo{journal}{\prb} \textbf{\bibinfo{volume}{81}}, \bibinfo{eid}{134509}
  (\bibinfo{year}{2010}), \eprint{0904.2197}.

\bibitem[{\citenamefont{{Potter} et~al.}(2016)\citenamefont{{Potter},
  {Morimoto}, and {Vishwanath}}}]{Potter:2016uq}
\bibinfo{author}{\bibfnamefont{A.~C.} \bibnamefont{{Potter}}},
  \bibinfo{author}{\bibfnamefont{T.}~\bibnamefont{{Morimoto}}},
  \bibnamefont{and}
  \bibinfo{author}{\bibfnamefont{A.}~\bibnamefont{{Vishwanath}}},
  \bibinfo{journal}{ArXiv e-prints}  (\bibinfo{year}{2016}),
  \eprint{1602.05194}.

\bibitem[{\citenamefont{{Maldacena} and {Stanford}}(2016)}]{Maldacena:2016tg}
\bibinfo{author}{\bibfnamefont{J.}~\bibnamefont{{Maldacena}}} \bibnamefont{and}
  \bibinfo{author}{\bibfnamefont{D.}~\bibnamefont{{Stanford}}},
  \bibinfo{journal}{ArXiv e-prints}  (\bibinfo{year}{2016}),
  \eprint{1604.07818}.

\bibitem[{\citenamefont{{Polchinski} and
  {Rosenhaus}}(2016)}]{Polchinski:2016qq}
\bibinfo{author}{\bibfnamefont{J.}~\bibnamefont{{Polchinski}}}
  \bibnamefont{and}
  \bibinfo{author}{\bibfnamefont{V.}~\bibnamefont{{Rosenhaus}}},
  \bibinfo{journal}{Journal of High Energy Physics}
  \textbf{\bibinfo{volume}{4}}, \bibinfo{eid}{1} (\bibinfo{year}{2016}),
  \eprint{1601.06768}.

\bibitem[{\citenamefont{{Gu} et~al.}(2016)\citenamefont{{Gu}, {Qi}, and
  {Stanford}}}]{Gu:2016wq}
\bibinfo{author}{\bibfnamefont{Y.}~\bibnamefont{{Gu}}},
  \bibinfo{author}{\bibfnamefont{X.-L.} \bibnamefont{{Qi}}}, \bibnamefont{and}
  \bibinfo{author}{\bibfnamefont{D.}~\bibnamefont{{Stanford}}},
  \bibinfo{journal}{ArXiv e-prints}  (\bibinfo{year}{2016}),
  \eprint{1609.07832}.

\bibitem[{\citenamefont{{Banerjee} and {Altman}}(2016)}]{Banerjee:2016dk}
\bibinfo{author}{\bibfnamefont{S.}~\bibnamefont{{Banerjee}}} \bibnamefont{and}
  \bibinfo{author}{\bibfnamefont{E.}~\bibnamefont{{Altman}}},
  \bibinfo{journal}{ArXiv e-prints}  (\bibinfo{year}{2016}),
  \eprint{1610.04619}.

\bibitem[{\citenamefont{{Witten}}(2016)}]{Witten:2016pb}
\bibinfo{author}{\bibfnamefont{E.}~\bibnamefont{{Witten}}},
  \bibinfo{journal}{ArXiv e-prints}  (\bibinfo{year}{2016}),
  \eprint{1610.09758}.

\bibitem[{\citenamefont{{Fu} et~al.}(2016)\citenamefont{{Fu}, {Gaiotto},
  {Maldacena}, and {Sachdev}}}]{Fu:2016la}
\bibinfo{author}{\bibfnamefont{W.}~\bibnamefont{{Fu}}},
  \bibinfo{author}{\bibfnamefont{D.}~\bibnamefont{{Gaiotto}}},
  \bibinfo{author}{\bibfnamefont{J.}~\bibnamefont{{Maldacena}}},
  \bibnamefont{and}
  \bibinfo{author}{\bibfnamefont{S.}~\bibnamefont{{Sachdev}}},
  \bibinfo{journal}{ArXiv e-prints}  (\bibinfo{year}{2016}),
  \eprint{1610.08917}.

\bibitem[{\citenamefont{{Berkooz} et~al.}(2016)\citenamefont{{Berkooz},
  {Narayan}, {Rozali}, and {Sim{\'o}n}}}]{Berkooz:2016th}
\bibinfo{author}{\bibfnamefont{M.}~\bibnamefont{{Berkooz}}},
  \bibinfo{author}{\bibfnamefont{P.}~\bibnamefont{{Narayan}}},
  \bibinfo{author}{\bibfnamefont{M.}~\bibnamefont{{Rozali}}}, \bibnamefont{and}
  \bibinfo{author}{\bibfnamefont{J.}~\bibnamefont{{Sim{\'o}n}}},
  \bibinfo{journal}{ArXiv e-prints}  (\bibinfo{year}{2016}),
  \eprint{1610.02422}.

\bibitem[{\citenamefont{{Gross} and {Rosenhaus}}(2016)}]{Gross:2016fy}
\bibinfo{author}{\bibfnamefont{D.~J.} \bibnamefont{{Gross}}} \bibnamefont{and}
  \bibinfo{author}{\bibfnamefont{V.}~\bibnamefont{{Rosenhaus}}},
  \bibinfo{journal}{ArXiv e-prints}  (\bibinfo{year}{2016}),
  \eprint{1610.01569}.

\bibitem[{\citenamefont{{You} and {Xu}}(2014)}]{You:2014pt}
\bibinfo{author}{\bibfnamefont{Y.-Z.} \bibnamefont{{You}}} \bibnamefont{and}
  \bibinfo{author}{\bibfnamefont{C.}~\bibnamefont{{Xu}}},
  \bibinfo{journal}{\prb} \textbf{\bibinfo{volume}{90}}, \bibinfo{eid}{245120}
  (\bibinfo{year}{2014}), \eprint{1409.0168}.

\bibitem[{\citenamefont{You et~al.}(2015)\citenamefont{You, Bi, Rasmussen,
  Cheng, and Xu}}]{You:2015bx}
\bibinfo{author}{\bibfnamefont{Y.-Z.} \bibnamefont{You}},
  \bibinfo{author}{\bibfnamefont{Z.}~\bibnamefont{Bi}},
  \bibinfo{author}{\bibfnamefont{A.}~\bibnamefont{Rasmussen}},
  \bibinfo{author}{\bibfnamefont{M.}~\bibnamefont{Cheng}}, \bibnamefont{and}
  \bibinfo{author}{\bibfnamefont{C.}~\bibnamefont{Xu}}, \bibinfo{journal}{New
  Journal of Physics} \textbf{\bibinfo{volume}{17}}, \bibinfo{pages}{075010}
  (\bibinfo{year}{2015}).

\bibitem[{\citenamefont{{Fidkowski} et~al.}(2013)\citenamefont{{Fidkowski},
  {Chen}, and {Vishwanath}}}]{Fidkowski:2013fk}
\bibinfo{author}{\bibfnamefont{L.}~\bibnamefont{{Fidkowski}}},
  \bibinfo{author}{\bibfnamefont{X.}~\bibnamefont{{Chen}}}, \bibnamefont{and}
  \bibinfo{author}{\bibfnamefont{A.}~\bibnamefont{{Vishwanath}}},
  \bibinfo{journal}{Physical Review X} \textbf{\bibinfo{volume}{3}},
  \bibinfo{eid}{041016} (\bibinfo{year}{2013}), \eprint{1305.5851}.

\bibitem[{\citenamefont{{Metlitski} et~al.}(2014)\citenamefont{{Metlitski},
  {Fidkowski}, {Chen}, and {Vishwanath}}}]{Metlitski:2014fv}
\bibinfo{author}{\bibfnamefont{M.~A.} \bibnamefont{{Metlitski}}},
  \bibinfo{author}{\bibfnamefont{L.}~\bibnamefont{{Fidkowski}}},
  \bibinfo{author}{\bibfnamefont{X.}~\bibnamefont{{Chen}}}, \bibnamefont{and}
  \bibinfo{author}{\bibfnamefont{A.}~\bibnamefont{{Vishwanath}}},
  \bibinfo{journal}{ArXiv e-prints}  (\bibinfo{year}{2014}),
  \eprint{1406.3032}.

\bibitem[{Sup()}]{SuppII}
\bibinfo{note}{{The meaning of the square of the time-reversal operator,
  $\mathcal{T}^2$, when acting on the many-body Fock space, is described in the
  Appendix \ref{sec: T^2}. Here we just note that, as shown in the Appendix,
  the time-reversal operator may acquire an additional phase, which can only be
  a fourth root of unity, when acting on the Fock space.}}

\bibitem[{\citenamefont{Santos and Rigol}(2010)}]{Rigol:2010ls}
\bibinfo{author}{\bibfnamefont{L.~F.} \bibnamefont{Santos}} \bibnamefont{and}
  \bibinfo{author}{\bibfnamefont{M.}~\bibnamefont{Rigol}},
  \bibinfo{journal}{Phys. Rev. E} \textbf{\bibinfo{volume}{81}},
  \bibinfo{pages}{036206} (\bibinfo{year}{2010}).

\bibitem[{\citenamefont{{Khemani} et~al.}(2014)\citenamefont{{Khemani},
  {Chandran}, {Kim}, and {Sondhi}}}]{Chandran:2014ls}
\bibinfo{author}{\bibfnamefont{V.}~\bibnamefont{{Khemani}}},
  \bibinfo{author}{\bibfnamefont{A.}~\bibnamefont{{Chandran}}},
  \bibinfo{author}{\bibfnamefont{H.}~\bibnamefont{{Kim}}}, \bibnamefont{and}
  \bibinfo{author}{\bibfnamefont{S.~L.} \bibnamefont{{Sondhi}}},
  \bibinfo{journal}{\pre} \textbf{\bibinfo{volume}{90}}, \bibinfo{eid}{052133}
  (\bibinfo{year}{2014}), \eprint{1406.4863}.

\bibitem[{\citenamefont{{Bar Lev} et~al.}(2015)\citenamefont{{Bar Lev},
  {Cohen}, and {Reichman}}}]{Reichman:2015sf}
\bibinfo{author}{\bibfnamefont{Y.}~\bibnamefont{{Bar Lev}}},
  \bibinfo{author}{\bibfnamefont{G.}~\bibnamefont{{Cohen}}}, \bibnamefont{and}
  \bibinfo{author}{\bibfnamefont{D.~R.} \bibnamefont{{Reichman}}},
  \bibinfo{journal}{Physical Review Letters} \textbf{\bibinfo{volume}{114}},
  \bibinfo{eid}{100601} (\bibinfo{year}{2015}), \eprint{1407.7535}.

\bibitem[{\citenamefont{{Atas} et~al.}(2013)\citenamefont{{Atas}, {Bogomolny},
  {Giraud}, and {Roux}}}]{Atas:2013la}
\bibinfo{author}{\bibfnamefont{Y.~Y.} \bibnamefont{{Atas}}},
  \bibinfo{author}{\bibfnamefont{E.}~\bibnamefont{{Bogomolny}}},
  \bibinfo{author}{\bibfnamefont{O.}~\bibnamefont{{Giraud}}}, \bibnamefont{and}
  \bibinfo{author}{\bibfnamefont{G.}~\bibnamefont{{Roux}}},
  \bibinfo{journal}{Physical Review Letters} \textbf{\bibinfo{volume}{110}},
  \bibinfo{eid}{084101} (\bibinfo{year}{2013}), \eprint{1212.5611}.

\bibitem[{\citenamefont{{Fu} and {Sachdev}}(2016)}]{Sachdev:ly}
\bibinfo{author}{\bibfnamefont{W.}~\bibnamefont{{Fu}}} \bibnamefont{and}
  \bibinfo{author}{\bibfnamefont{S.}~\bibnamefont{{Sachdev}}},
  \bibinfo{journal}{ArXiv e-prints}  (\bibinfo{year}{2016}),
  \eprint{1603.05246}.

\bibitem[{\citenamefont{{Wang} and {Senthil}}(2014)}]{Senthil:2014qy}
\bibinfo{author}{\bibfnamefont{C.}~\bibnamefont{{Wang}}} \bibnamefont{and}
  \bibinfo{author}{\bibfnamefont{T.}~\bibnamefont{{Senthil}}},
  \bibinfo{journal}{\prb} \textbf{\bibinfo{volume}{89}}, \bibinfo{eid}{195124}
  (\bibinfo{year}{2014}), \eprint{1401.1142}.

\bibitem[{\citenamefont{Chiu et~al.}(2016)\citenamefont{Chiu, Teo, Schnyder,
  and Ryu}}]{ChiuTeoSchnyderRyuRMP2016}
\bibinfo{author}{\bibfnamefont{C.}~\bibnamefont{Chiu}},
  \bibinfo{author}{\bibfnamefont{J.}~\bibnamefont{Teo}},
  \bibinfo{author}{\bibfnamefont{A.}~\bibnamefont{Schnyder}}, \bibnamefont{and}
  \bibinfo{author}{\bibfnamefont{S.}~\bibnamefont{Ryu}}, \bibinfo{journal}{Rev.
  Mod. Phys.} \textbf{\bibinfo{volume}{88}}, \bibinfo{pages}{035005}
  (\bibinfo{year}{2016}).

\end{thebibliography}

\onecolumngrid
\newpage
\begin{center}
\textbf{\large Appendix}
\end{center}
\vspace{0.5cm}
\setcounter{equation}{0}
\setcounter{figure}{0}
\setcounter{table}{0}
\setcounter{page}{1}
\makeatletter
\renewcommand{\theequation}{S\arabic{equation}}
\renewcommand{\thetable}{S\Roman{table}}
\renewcommand{\thefigure}{S\arabic{figure}}
\twocolumngrid
\section{Projective Representation Analysis for Symmetry Class BDI}\label{sec: BDI}
In the main text, we have shown that for symmetry classes AIII and CII, the (projective) symmetry action on the boundary restricts the boundary Hamiltonian $H$ to either real, complex or quaternion Hermitian matrices, and hence exhibiting the three classes of Wigner-Dyson level statistics. In this appendix, we will show that the same kind of argument can be applied to  symmetry class BDI as well, which will provide another perspective to understand the level statistics apart from the Clifford algebra argument given in the main text.

The projective symmetry representation
on the many-body Hilbert space at the boundary of a 1D Fermion system in symmetry class
BDI case has been thoroughly studied by Fidkowski and Kitaev in their pioneering work Ref.\,\cite{Fidkowski:2011pa}. Here we will briefly review some results of Ref.\,\cite{Fidkowski:2011pa}, and then discuss the their  implications on the level statisics. For the Fermion chain
in symmetry class  BDI,
the full symmetry group in consideration is $Z_2^P\times Z_2^T$, where $Z_2^P$ is the Fermion parity symmetry and $Z_2^T$ is the time-reversal symmetry. The many-body state of the boundary Majorana modes form a projective representation of this symmetry group. 

In terms of the Majorana operators $\chi_a$ ($a=1,2,\cdots,N_\chi$) on the boundary, the Fermion parity operator $P$ can be written as
\eq{\label{eq: P BDI}P=\left\{\begin{array}{ll} (-\ii\chi_1\chi_2)(-\ii\chi_3\chi_4)\cdots(-\ii\chi_{N_\chi-1}\chi_{N_\chi}) & N_\chi\in\text{even,}\\(-\ii\chi_1\chi_2)(-\ii\chi_3\chi_4)\cdots(-\ii\chi_{N_\chi}\chi_\infty) & N_\chi\in\text{odd,}\end{array}\right.}
such that the Fermion parity operator anti-commutes with all Majorana Fermion operators, i.e. $\forall a: \chi_a P = -P \chi_a$, as expected. Note that for odd $N_\chi$, an extra Majorana mode $\chi_\infty$ at infinity is added to complete the physical Hilbert space, and also to make $P$ operator itself an even-Fermion-parity operator. As shown in Ref.\,\cite{Fidkowski:2011pa}, there is no non-trivial projective representation associated to $P^2$, meaning that one can always make $P^2=1$ by gauge fixing, and such a gauge choice has been made in \eqnref{eq: P BDI}.

For odd $N_\chi$, Ref.\,\cite{Fidkowski:2011pa} also introduces an useful operator $Z$ by ``factoring out'' the extra Majorana Fermion $\chi_\infty$ from the Fermion parity operator $P$, so that 
 $P=\ii\chi_\infty Z$. One can see that  $Z$ is similar to $P$ but does not involve $\chi_\infty$,
\eq{\label{eq: Z BDI}Z=(-\ii)^{(N_\chi-1)/2}\prod_{a=1}^{N_\chi}\chi_a \quad (N_\chi\in\text{odd}).}
$Z$ squares to one (i.e. $Z^2=1$) and anti-commutes with $P$ (i.e. $ZP=-PZ$). Importantly, $Z$ commutes with all Fermion interaction terms 
(which are sum of products of four $\chi_a$ operators),
 and thus $Z$ also commutes with the boundary Hamiltonian $H$. So $Z$ is an additional symmetry of the Hamiltonian $H$ in the case of odd $N_\chi$. 

As an anti-unitary operator, the time-reversal operator $\mathcal{T}=\mathcal{U}_T  \mathcal{K}$ can be considered as complex conjugation $\mathcal{K}$ followed by a unitary transformation $\mathcal{U}_T$. 
One needs to
specify the meaning of $\mathcal{K}$
(which is basis-dependent)
 as follows (following Ref.\,\cite{Fidkowski:2011pa}). First we pick a Fermion occupation number basis (Fock basis) by assuming that the complex Fermion annihilation and creation operators $c_m$ and $c_m^\dagger$ (for $m=1,2,\cdots$) are defined as 
\eq{c_m=\tfrac{1}{2}(\chi_{2m-1}+\ii\chi_{2m}),\quad c_m^\dagger=\tfrac{1}{2}(\chi_{2m-1}-\ii\chi_{2m}).}
For odd $N_\chi$, we will include $\chi_\infty$ to define the last pair of complex Fermion operators. Let $\ket{0}$ be the state annihilated by all the $c_m$ operators. Any Fermion many-body state $\ket{\psi}$ in the boundary Hilbert space can be represented in the Fock basis as
\eq{\ket{\psi}=\sum_{n_m\in\{0,1\}}C_{n_1n_2\cdots}{c_1^\dagger}^{n_1}{c_2^\dagger}^{n_2}\cdots\ket{0}.}
Now we define $\mathcal{K}$ to be the complex conjugation operator in this basis of Fock space, which leaves the basis kets
 ${c_1^\dagger}^{n_1}{c_2^\dagger}^{n_2}\cdots\ket{0}$
invariant and
acts by complex conjugating the coefficients $C_{n_1n_2\cdots}$.
With this definition of complex conjugation,
the Majorana Fermion operator $\chi_a$ will have an alternating sign under complex conjugation depending on whether the index
$a$ is even or odd, i.e. $\mathcal{K}\chi_{a}\mathcal{K}=-(-1)^a\chi_{a}$. This alternating sign must be compensated for 
by the unitary transformation $\mathcal{U}_T$ via $\mathcal{U}_T \chi_a \mathcal{U}_T^{-1}=-(-1)^a\chi_a$,  so that the time-reversal transformation $\mathcal{T}\chi_a\mathcal{T}^{-1}=\chi_a$ leaves the Fermion operator $\chi_a$ unchanged.
A unitary operator satisfying this condition
\eq{\label{eq: T BDI}
\mathcal{U}_T=P^{\lceil N_\chi/2\rceil +1}\prod_{a=1:2:N_\chi}\chi_{a},}
where $\lceil N_\chi/2\rceil$ denotes the smallest integer larger than  
$N_\chi/2$ (``integer ceiling''), and $a=1:2:N_\chi$ means that $a$ steps from 1 to $N_\chi$ with increment 2. When $N_\chi$ is odd, there is an ambiguity in the choice of $\mathcal{U}_T$, because the transform of $\chi_\infty$ under $\mathcal{T}$ is not specified. 
In this case we choose $\mathcal{T}\chi_\infty\mathcal{T}^{-1}=\chi_\infty$, which differs from the choice made in Ref.\,\cite{Fidkowski:2011pa} for $N_\chi=$1 and 5. However, with our choice, the time-reversal operator 
has a unified expression,  \eqnref{eq: T BDI}, for all $N_\chi$.

Using the explicit representations for $P$ in \eqnref{eq: P BDI}, for  $Z$ in \eqnref{eq: Z BDI}, and 
for $\mathcal{T}=\mathcal{U}_T\mathcal{K}$ in \eqnref{eq: T BDI}, their algebraic relations can be explicitly calculated, and the result is 
summarized
in \tabref{tab: symmetry BDI}. The projective representations are fully classified by three invariants: $\mathcal{T}^2$, $(P\mathcal{T})^2$ and $(Z\mathcal{T})^2$, where the last one $(Z\mathcal{T})^2$ is only defined for odd $N_\chi$. In particular, $(P\mathcal{T})^2$ distinguishes the topological index $\nu$ from $-\nu$ (where $\nu\equiv N_\chi(\mod 8)$). So by combining the invariant $(P\mathcal{T})^2$ with the level statistics, one can fully resolve the $\dsZ_8$ anomaly pattern.

\begin{table}[htbp]
\caption{Projective symmetry group invariants that distinguish the $\dsZ_8$ anomaly.}
\begin{center}
\begin{tabular}{c|cccccccc}
$N_\chi(\mod 8)$ & 0 & 1 & 2 & 3 & 4 & 5 & 6 & 7\\
\hline
$\mathcal{T}^2$ & $+$ & $+$ & $+$ & $-$ & $-$ & $-$ & $-$ & $+$ \\
$(P\mathcal{T})^2$ & $+$ & $-$ & $-$ & $-$ & $-$ & $+$ & $+$ & $+$ \\
$(Z\mathcal{T})^2$ & & $+$ & & $+$ & & $-$ & & $-$ 
\end{tabular}
\end{center}
\label{tab: symmetry BDI}
\end{table}

Having determined the algebraic relations between $P$, $Z$ and $\mathcal{T}$, we seek the 
 matrix representations of the symmetries and the Hamiltonian in the boundary many-body Hilbert space for all $N_\chi ({\rm mod}~8)$. 
We can work in the block-diagonal basis of the Fermion parity operator $P$, so that
\eq{P=\mat{\mathbf{1} & 0\\ 0& -\mathbf{1}}, \ {\rm and} \  Z=\mat{0&\mathbf{1} \\ \mathbf{1}&0}\text{ (for odd $N_\chi$)},}
satisfy $P^2=Z^2=1$ and $ZP = - PZ$. In this basis, to meet the requirements of $\mathcal{T}^2$, $(P\mathcal{T})^2$ and $(Z\mathcal{T})^2$ listed in \tabref{tab: symmetry BDI}, the representations of the time-reversal operator $\mathcal{T}=\mathcal{U}_T\mathcal{K}$ can be determined, as 
summarized  in \tabref{tab: rep BDI}. 
Here,  $\Omega$ is a real matrix that squares to $-1$, i.e. $\Omega^2=-1$. Without 
loss of generality, we may choose $\Omega$ to be 
$\Omega=\mat{0 & +\mathbf{1}\\-\mathbf{1} & 0}$.

\begin{table}[htbp]
\caption{Representations of $\mathcal{U}_T$ and $H$ that are consistent with all the algebraic relations. Here, $\Omega$ can be any real matrix that squares to $\Omega^2=-1$. $H_\dsR$, $H_\dsC$, $H_\dsH$ stands for real, complex and quaternion Hermitian matrices. A prime on $H'$ indicates $H'$ is in general differed from $H$. The Clifford algebra $\Cl_{0, N_\chi-1}$ and level statistics (lev. stat.) in each Fermion number parity sector are also listed.}
\begin{center}
\begin{tabular}{c|cccc}
$N_\chi(\mod 8)$ & 0 & 1 & 2 & 3\\
\hline
$\mathcal{U}_T=$ &
$\mat{\mathbf{1} & 0\\ 0& -\mathbf{1}}$ &
$\mat{0 & \mathbf{1}\\ \mathbf{1} & 0}$ &
$\mat{0 & \mathbf{1}\\ \mathbf{1} & 0}$ &
$\mat{\Omega & 0\\ 0& -\Omega}$ \\
$H=$ &
$\mat{H_{\dsR} & 0\\ 0 & H'_{\dsR}}$ &
$\mat{H_{\dsR} & 0\\ 0 & H_{\dsR}}$ &
$\mat{H_{\dsC} & 0\\ 0 & H^*_{\dsC}}$ &
$\mat{H_{\dsH} & 0\\ 0 & H_{\dsH}}$ \\
$\Cl_{0, N_\chi-1}$ &
$\dsR\oplus\dsR$ &
$\dsR$ &
$\dsC$ &
$\dsH$ \\
lev. stat. & GOE & GOE & GUE & GSE \\
\hline\hline
$N_\chi(\mod 8)$ & 4 & 5 & 6 & 7\\
\hline
$\mathcal{U}_T=$ &
$\mat{\Omega & 0\\ 0& -\Omega}$ &
$\mat{0 & \Omega\\ \Omega & 0}$ &
$\mat{0 & \mathbf{1}\\ -\mathbf{1} & 0}$ &
$\mat{\mathbf{1} & 0\\ 0& -\mathbf{1}}$ \\
$H=$ &
$\mat{H_{\dsH} & 0\\ 0 & H'_{\dsH}}$ &
$\mat{H_{\dsH} & 0\\ 0 & H_{\dsH}}$ &
$\mat{H_{\dsC} & 0\\ 0 & H^*_{\dsC}}$ &
$\mat{H_{\dsR} & 0\\ 0 & H_{\dsR}}$ \\
$\Cl_{0, N_\chi-1}$ &
$\dsH\oplus\dsH$ &
$\dsH$ &
$\dsC$ &
$\dsR$ \\
lev. stat. & GSE & GSE & GUE & GOE \\
\end{tabular}
\end{center}
\label{tab: rep BDI}
\end{table}

The Hamiltonian $H$ must respect all the symmetries. From the Fermion parity symmetry $PH=HP$, we know $H$ must be block diagonal, and takes the form of
\eq{H=\mat{H_+ & 0 \\ 0 & H_-},}
where $H_+$ ($H_-$) is the Hamiltonian that acts in the even (odd) Fermion parity subspace. If the number
$N_\chi$ of  Majorana  operators
is odd, $Z$ is an additional symmetry of $H$. Then $ZH=HZ$ further requires $H_+ = H_-$ for odd $N_\chi$. Finally $\mathcal{T}H=H\mathcal{T}$ implies $\mathcal{U}_T H^* \mathcal{U}_T^{-1} = H$. Using the representation
 of $\mathcal{U}_T$ listed in \tabref{tab: rep BDI}, we can determine whether $H_+$ and $H_-$ are matrices
with  real, complex or quaternion matrix elements, and the result is 
summarized in \tabref{tab: rep BDI}. $H_\dsR$, $H_\dsC$, and $H_\dsH$ stand for the set of
$n\times n$ Hermitian matrices with real, complex and quaternion matrix elements with some  $n$.
The  prime on $H'$ indicates that  $H'$ is in general differed from $H$. One can see that the result is consistent with the Clifford algebra analysis (by considering $\Cl_{0,N_\chi-1}$)
discussed in the 
main
text. So we reach at the same conclusion about the level statistics from the analysis of the projective symmetry representations
on
the many-body Hilbert space at the boundary.

\section{Square of Anti-Unitary Symmetries in Many-Body Fock Space}\label{sec: T^2}

In this Appendix we discuss in general the action of an anti-unitary operator
$\Theta$ such as the  time-reversal $\Theta=\mathcal{T}$ or
the  chiral symmetry  operation $\Theta = \mathcal{S}$ on the many-body Fock space
of a system of Fermions. The square of these operators, as defined by its action on
the Fermion creation and annihilation operators (which determine   their action on the single-particle Hilbert space)
is characterized in the familiar way by a number that we call $\gamma_\text{sp}$, which can take only values  
$\gamma_\text{sp}=\pm 1$  (``single-particle phase'') - see Eq.s (\ref{eq: T^2 SP},\ref{LabelEqDEFTimeReversalSquareComplexFermions},\ref{LabelEqDEFChiralSquareComplexFermions}).
Here we show that the action of the square of the same anti-unitary operators on the many-body Fock space many acquire
an additional {\it many-body phase} $\gamma_\text{mb}$, see  \eqnref{LabelEqThetaSquareManyBody}, whose value is related to $\gamma_\text{sp}$
in the manner displayed in \eqnref{eq: dep}. This phase can only be a 4th root on unity. The notion of $\mathcal{T}^2=\pm\ii$ was also discussed in Ref.\,\onlinecite{Fidkowski:2013fk,Metlitski:2014fv}.


Consider first  a many-body system defined by a set of Majorana Fermion operators $\chi_j = \chi^\dagger_j$, where
 $\{\chi_i,\chi_j\}=2\delta_{ij}$. Let $\mathcal{T}$ be the time-reversal operator. 
The meaning of the time-reversal operation at the single-particle level is defined by its action on the canonical Majorana Fermion operators,
\begin{equation}
\label{LabelEqDEFMajoranaTimeReversal}
\mathcal{T} \chi_j  \mathcal{T}^{-1} = \sum_k  W_{jk} \chi_k
\end{equation} 
where in order to preserve the canonical anti-commutation relations,  $W_{jk}$ is an orthogonal matrix.
At the single-particle level, the meaning of the square of the time-reversal operator,
``$\mathcal{T}^2=\gamma_\text{sp}=\pm1$'', 
is defined by its  action on the canonical Fermion operators,
\eq{\label{eq: T^2 SP}
 \mathcal{T}^2\chi_i\mathcal{T}^{-2}=\gamma_\text{sp}\chi_i,
\ \ (i=1, 2, ...)}
where the sign factor 
$\gamma_\text{sp}=
\pm1$ characterizes the square of the ``{\it single-particle}'' time-reversal operator. 
Here
\begin{equation}
\label{LabelEqSingleParticleTimeReversalSquaredMajorana}
\sum_j
W_{ij} W_{jk} = \gamma_\text{sp} \delta_{ik}.
\end{equation}
Consider now the time-reversal operator $\mathcal{T}$  when acting on the
many-body Hilbert space (Fock space).  As an anti-unitary operator, it takes on general grounds  
the form 
\eq{\mathcal{T}=
{\cal U}_T
\mathcal{K},}
where ${\cal U}_T$ is a unitary operator acting on the many-body Hilbert space, and $\mathcal{K}$ denotes the complex conjugation operator acting on the same space. 
As a consequence of \eqnref{eq: T^2 SP}
the 
square of the time-reversal operator acting on the many-body Fock space
 takes in general  the form
\eq{\label{eq: T^2 MB}
\mathcal{T}^2={\cal U}_T \mathcal{K} {\cal U}_T \mathcal{K}=
{\cal U}_T {\cal U}_T^*
=\gamma_\text{mb}(\gamma_\text{sp})^F,}
where $\gamma_\text{sp}^F=(\pm1)^F$ is the 
 Fermion number parity operator when 
$\gamma_\text{sp}=-1$.
The point we want to stress in this Appendix is
that  there can be  an {\it extra phase} 
$\gamma_\text{mb}$ (``many-body phase'')  that cannot be removed, or  ``gauged away''.
\eqnref{eq: T^2 MB}, {\it containing} this  additional phase
 $\gamma_\text{mb}$, defines the notion of the many-body
time-reversal operator $\mathcal{T}^2$ that is used throughout this paper.
 As a consistency check, one
immediately sees that \eqnref{eq: T^2 MB} is consistent with \eqnref{eq: T^2 SP}, since
$\mathcal{T}^2\chi_i\mathcal{T}^{-2}=\gamma_\text{mb}\gamma_\text{sp}^F\chi_i\gamma_\text{sp}^{-F}\gamma_\text{mb}^{-1}=\gamma_\text{sp}\chi_i$.
The many-body phase
 $\gamma_\text{mb}$ always cancels out  in this equation as $\gamma_\text{mb}\gamma_\text{mb}^{-1}=1$.

Moreover,  an expression of the form of \eqnref{eq: T^2 MB} holds true in general for both\cite{Ludwig:2010xt,Ludwig:2015la}
 {\it anti-unitary} operators
in Fock space, the time-reversal operator $\mathcal{T}$ as well as the chiral symmetry operator ${\cal S}$. The former
acts on canonical Fermion creation- and annihilation operators as
\begin{equation}
\label{LabelEqDEFTimeReversalComplexFermions}
\mathcal{T} c^\dagger_j \mathcal{T}^{-1} = \sum_k c^\dagger_k \ U_{kj};
\ \ \ 
\mathcal{T} c_j \mathcal{T}^{-1} =\sum_k (U^\dagger)_{j,k}  c_k
\end{equation}
where $U$ is a unitary matrix
and
\begin{equation}
\label{LabelEqDEFTimeReversalSquareComplexFermions}
\mathcal{T}^2 c_j \mathcal{T}^{-2} = \gamma_{\text{sp}} c_j,
\end{equation}
with $\gamma_{\text{sp}}=\pm 1$.
The chiral symmetry  acts on the same operators as
\begin{equation}
\label{LabelEqDEFChiralComplexFermions}
\mathcal{S} c^\dagger_j \mathcal{S}^{-1} = \sum_k c_k \ V_{kj};
\ \ \ 
\mathcal{S} c_j \mathcal{S}^{-1} \sum_k (V^\dagger)_{j,k}  c^\dagger_k
\end{equation}
where $V$ is  a unitary matrix and
\begin{equation}
\label{LabelEqDEFChiralSquareComplexFermions}
\mathcal{S}^2 c_j \mathcal{S}^{-2} = \gamma_{\text{sp}} c_j.
\end{equation}
Here (for the chiral symmetry) it is always possible\cite{ChiuTeoSchnyderRyuRMP2016,Ludwig:2015la}
 to choose  
$\gamma_{\text{sp}}=1$.

If we now denote a general {\it anti-unitary} operator in the many-body Fock space by $\Theta$,
representing either time-reversal, $\Theta = \mathcal{T}$, or chiral symmetry, $\Theta = \mathcal{S}$, then,
owing to Eq.s\,(\ref{LabelEqDEFTimeReversalSquareComplexFermions},\ref{LabelEqDEFChiralSquareComplexFermions}),
its square has
in general the form
\begin{equation}
\label{LabelEqThetaSquareManyBody}
\Theta^2 =
\gamma_\text{mb} \ 
( \gamma_{\text{sp}})^F 
\end{equation}
where $\gamma_\text{mb}$ is a phase.
We will now demonstrate that in this general setting the possible choices
for the phase  $\gamma_\text{mb}$ are  related to the value of $\gamma_\text{sp}$ in the following way:
\eq{\label{eq: dep}
\left\{\begin{array}{ll}
\gamma_\text{mb}=+1, -1 & \text{if }\gamma_\text{sp}=+1;\\
\gamma_\text{mb}=\pm1, \pm\ii & \text{if }\gamma_\text{sp}=-1.\\
\end{array}\right.}
Before proving \eqnref{eq: dep} let us list the
following examples of this result that  apply to systems discussed in this paper: (i): For the time-reversal operator
$\Theta=\mathcal{T}$ in  symmetry class BDI, for which  $\gamma_\text{sp}=+1$,
its square in Fock space is  $\mathcal{T}^2 = \gamma_\text{mb} {\bf 1}$ with $\gamma_\text{mb}=+1$
or $\gamma_\text{mb}=-1$. (See also Table \ref{tab: symmetry BDI} of the Appendix.) (ii): For the Chiral symmetry operator
$\Theta=\mathcal{S}$ in symmetry class AIII, for which we choose by convention $\gamma_\text{sp}=+1$,
its square  in Fock space can be $\mathcal{S}^2 =\gamma_\text{mb} {\bf 1}$ with $\gamma_\text{mb}=+1$
or $\gamma_\text{mb}=-1$.  (See 
\eqnref{eq:ChiralSymmetrySquaredFockSpace} 
in the main text.)
(iii): For the time-reversal operator $\Theta=\mathcal{T}$ in symmetry class CII, which has 
$\gamma_\text{sp}=-1$, 
 its square
in Fock space takes on values $\mathcal{T}^2 = \gamma_\text{mb} (-1)^F$ with $\gamma_\text{mb} =(-1)^{N_f}$ in the examples
given in the main part of this paper.

Let us now proceed to the proof of \eqnref{eq: dep}.
To this end, 
consider
\begin{eqnarray}
\label{LabelEqThetaCubed}
&&\Theta^3 = \Theta^2 \  \Theta = \Theta \  \Theta^2=
\\ \nonumber
&&
=\gamma_\text{mb}  \  (\gamma_{\text{sp}})^F \  \Theta  =  \Theta \ \gamma_\text{mb}  \  (\gamma_{\text{sp}})^F
\end{eqnarray}
where in the last line use was made of
\eqnref{LabelEqThetaSquareManyBody}.
The anti-linearity of $\Theta$ implies $\Theta \gamma_\text{mb}= \gamma_\text{mb}^* \Theta
= \gamma_\text{mb}^{-1}\Theta
$ (since $\gamma_\text{mb}$ is a phase), so that \eqnref{LabelEqThetaCubed} leads to 
\begin{equation}
\label{LabelEqThetaCubed-Two}
 (\gamma_{\text{sp}})^F \  \Theta  \  (\gamma_{\text{sp}})^F 
 =  
(\gamma_\text{mb})^{-2} \ 
\Theta 
\end{equation}
where we have used  the fact that
$(\gamma_\text{sp}^F)^2=(\gamma_\text{sp}^2)^F=1$.
Now, if $\gamma_\text{sp}=+1$, \eqnref{LabelEqThetaCubed-Two}
becomes $\Theta=\gamma_\text{mb}^{-2}\Theta$, implying $\gamma_\text{mb}^2=1$, and hence
 $\gamma_\text{mb}$ can only be $\pm1$ when $\gamma_\text{sp}=+1$. On the other hand, if  $\gamma_\text{sp}=-1$, we can conjugate
both sides of \eqnref{LabelEqThetaCubed-Two}
with  $\gamma_\text{sp}^F$  which, using  again $(\gamma_\text{sp}^F)^2=(\gamma_\text{sp}^2)^F=1$, yields
\begin{equation}
\label{LabelEqThetaCubed-Two-Conjugate}
\Theta
 =  
(\gamma_\text{mb})^{-2} \ (\gamma_{\text{sp}})^F \  \Theta  \  (\gamma_{\text{sp}})^F 
=
(\gamma_\text{mb})^{-4} \ \  \Theta.
\end{equation}
Therefore we conlude that $\gamma_\text{mb}^4=1$, meaning that $\gamma_\text{mb}$ must be a 4th root of unity, 
and hence  can only take the values  $\pm1$ and $\pm\ii$.
In conclusion, we have demonstrated the  dependence of the many-body phase $\gamma_\text{mb}$ on the single-particle
sign $\gamma_\text{sp}$ (which determines the square of the anti-unitary operator at the single-particle level),
which was claimed  in \eqnref{eq: dep}.

\end{document}